\documentclass[10pt,a4paper]{amsart}
\usepackage{amssymb,amsmath}
\usepackage[american,french]{babel}
\usepackage{amsopn}
\usepackage{mathrsfs}
\usepackage{epstopdf}
\usepackage{graphicx}
\usepackage{url}
\usepackage[ec]{aeguill}
\usepackage{fancyhdr}
\usepackage[T1]{fontenc}
\newcommand{\rr}{\mathbb{R}}

\newcommand{\eps}{\varepsilon}
\newcommand{\nn}{\mathbb{N}}
\newcommand{\cc}{\mathbb{C}}

\newcommand{\R}{\mathbb{R}}

\def\wrtext#1{\relax\ifmmode{\leavevmode\hbox{#1}}\else{#1}\fi}
\def\abs#1{\left|#1\right|}
\def\begeq{\begin{equation}}
\def\endeq{\end{equation}}

\def\part#1{\frac{\partial}{\partial #1}}

\begin{document}

\newcounter{equa}
\selectlanguage{american}

\newtheorem{lemma}{Lemma}[section]
\newtheorem{definition}{Definition}[section]
\newtheorem{proposition}{Proposition}[section]
\newtheorem{theorem}{Theorem}[section]

\title[RETURN TO EQUILIBRIUM FOR HYPOELLIPTIC QUADRATIC SYSTEMS]{EXPONENTIAL RETURN TO EQUILIBRIUM FOR HYPOELLIPTIC QUADRATIC SYSTEMS}
\author{M. \textsc{Ottobre}, G.A. \textsc{Pavliotis} \& K. \textsc{Pravda-Starov}}
\address{\noindent \textsc{Department of Mathematics,
Imperial College London,
Huxley Building, 180 Queen's Gate, 
London SW7 2AZ, UK}
} 
\email{m.ottobre08@imperial.ac.uk, g.pavliotis@imperial.ac.uk}
\address{\noindent \textsc{Universit\'e de Cergy-Pontoise, CNRS UMR 8088, D\'epartement de Math\'ematiques,
95000 Cergy-Pontoise, France}
} 

\email{karel.pravda-starov@u-cergy.fr}

\begin{abstract} We study the problem of convergence to equilibrium for evolution equations associated to general quadratic operators.
Quadratic operators are non-selfadjoint differential operators with complex-valued quadratic symbols. Under appropriate assumptions, a complete description of the spectrum of such operators is given and the exponential return to equilibrium with sharp estimates on the rate of convergence is proven. Some applications to the study of chains of oscillators and the generalized Langevin equation are given.
\end{abstract}

\keywords{Quadratic operators, hypoellipticity, return to equilibrium, rate of convergence, chains of oscillators, generalized Langevin equation} 
\subjclass[2000]{Primary: 35H10, 35P99; Secondary: 35Q82}

\maketitle

\section{Introduction}
\label{sec:intro}

The spectral properties of hypoelliptic operators and the problem of return to equilibrium for the associated parabolic equations are topics of current interest. Such operators arise naturally in non-equilibrium statistical mechanics as the generators of Markov processes which model the dynamics of classical open systems~\cite{EH00, EckmPillR-B00, EPR99,  OttobrePavliotis09,Rey-Bellet2006}. A standard example of hypoelliptic equation is the Fokker-Planck equation~\cite{Ris84}, 
\begin{equation}\label{e:fp}
\left\lbrace
\begin{array}{l}
\partial_t f + v \cdot \nabla_x f - \nabla_x V(x) \cdot \nabla_v f -\gamma \nabla_v \cdot \left( v + \beta^{-1} \nabla_v \right) f  =  0, \\
            f(x,v,t)|_{t=0}  =  f_0(x,v).
\end{array}
\right.
\end{equation}
This equation is the forward Kolmogorov equation for the Langevin stochastic differential equation
\begin{equation}\label{e:langevin}
\left\lbrace
\begin{array}{l}
\dot{x} = v, \\
\dot{v} = -\nabla_x V(x) - \gamma v + \sqrt{2 \gamma \beta^{-1}} \dot{W},
\end{array}
\right.
\end{equation}
where $W(t)$ denotes a standard Brownian motion in $\R^d$, $\gamma$ the friction coefficient and $\beta$ the inverse temperature. Several properties of the Fokker-Planck equation have been studied including spectral properties and convergence to equilibrium~\cite{EH00, HelNie05HES, HerNie02IHA}, homogenization~\cite{HP07,HairPavl04}, short time asymptotics~\cite{Herau2007} and semiclassical limits~\cite{HeHiSj,HeHiSj2,HeHiSj3, HeSjSt}. Related problems have been studied for the Fokker-Planck equation corresponding to the generalized Langevin equation under a Markovian approximation~\cite{OttobrePavliotis09}.

Recently, the theory of hypocoercivity~\cite{Vil04HPI} has been developed for studying the convergence to equilibrium of kinetic equations. This theory applies to evolution equations of the form
\begin{equation}\label{e:hypocoer}
\partial_t f + (A^* A + X_0) f = 0, 
\end{equation}
where $A, \, X_0$ are some operators defined on an appropriate Hilbert space $\mathcal{H}$, which is usually the $L^2$ space weighted by the invariant measure of the underlying Markov process, with $X_0$ being skew-adjoint in this space, $X_0^* = - X_0$. Under appropriate assumptions on the commutators between the operators $A, \, X_0$ and higher order commutators, it is possible to prove the exponential return to equilibrium for these evolution equations. 
Regarding the Fokker-Planck operator, quantitative estimates on the rate of convergence can be obtained via the analysis of the spectral gap. However, it is not possible in general to determine exactly the spectrum and one of the few instances when this is possible is the case of quadratic potentials~\cite{HeSjSt}.

In the present work, we consider evolution equations associated with general quadratic operators
\begin{equation}\label{sm1}
\left\lbrace
\begin{array}{c}
\displaystyle \frac{\partial u}{\partial t}(t,x)+q^w(x,D_x) u(t,x)=0  \\
u(t,\textrm{\textperiodcentered})|_{t=0}=u_0 \in L^2(\rr^n),
\end{array} \right.
\end{equation}
and address the problem of the exponential return to equilibrium for these systems. 
Quadratic operators are pseudodifferential operators
defined in the Weyl quantization
\begin{equation}\label{3}
q^w(x,D_x) u(x) =\frac{1}{(2\pi)^n}\int_{\rr^{2n}}{e^{i(x-y).\xi}q\Big(\frac{x+y}{2},\xi\Big)u(y)dyd\xi}, 
\end{equation}
by symbols $q(x,\xi)$, with $(x,\xi) \in \rr^{n} \times \rr^n$ and $n \in \nn^*$, which
are complex-valued quadratic forms 
$$q : \rr_{x}^{n} \times \rr_{\xi}^{n} \rightarrow \cc.$$
These operators are differential operators with simple and fully explicit expression. Indeed, the Weyl quantization of the quadratic symbol
$x^{\alpha} \xi^{\beta}$, with $(\alpha,\beta) \in \nn^{2n}$, $|\alpha+\beta| = 2$, is the differential operator
$$
\frac{x^{\alpha}D_x^{\beta}+D_x^{\beta} x^{\alpha}}{2}, \ D_x=i^{-1}\partial_x.
$$
Notice that quadratic operators are non-selfadjoint operators and that those with symbols having non-negative real parts are accretive. Since the classical work by J.~Sj\"ostrand~\cite{sjostrand}, a complete description for the spectrum of elliptic quadratic operators is known and has played an important r\^ole in the analysis of partial differential operators with double characteristics. Elliptic quadratic operators are quadratic operators whose symbols satisfy the ellipticity condition
\begin{equation}\label{yo1}
(x,\xi) \in \rr^{2n}, \ q(x,\xi)=0 \Rightarrow (x,\xi)=0.
\end{equation} 
In a recent work~\cite{hps}, the spectral properties of non-elliptic quadratic operators, that is operators whose symbols may fail to satisfy the ellipticity condition on the whole phase space $\rr^{2n}$, were investigated. For any quadratic operator whose symbol has a real part with a sign, say here a symbol with non-negative real part 
\begin{equation}\label{smm1}
\textrm{Re }q \geq 0,
\end{equation}
it was pointed out the existence of a particular linear subvector space $S$ in the phase space $\rr^{2n}$, $S \subset \rr^{2n}$, intrinsically associated to the symbol $q$ and
called singular space, which plays a basic r\^ole in the understanding of the properties of this non-elliptic quadratic operator. In particular, this work~\cite{hps} (Theorem~1.2.2) gives a complete description for the spectrum of any non-elliptic quadratic operator $q^w(x,D_x)$ whose symbol $q$ has a non-negative real part, $\textrm{Re }q \geq 0$ and satisfies an assumption of
partial ellipticity along its singular space~$S$,
\begin{equation}\label{sm2}
(x,\xi) \in S, \ q(x,\xi)=0 \Rightarrow (x,\xi)=0. 
\end{equation}
Under those assumptions, the spectrum of the quadratic operator $q^w(x,D_x)$ is showed to be composed of a countable number of eigenvalues with finite multiplicity and the structure of the spectrum is 
similar to the one known for elliptic quadratic operators~\cite{sjostrand}. This condition of partial ellipticity is weaker than the condition of ellipticity in general, $S \subsetneq \rr^{2n}$, and allows to deal with more degenerate situations. An important class of quadratic operators satisfying condition (\ref{sm2})  are those with zero singular space $S=\{0\}$. In this case, the condition of partial ellipticity trivially holds and these differential operators have been showed to be hypoelliptic \cite{karel} (Theorem~1.2.1). More precisely, they enjoy specific subelliptic properties with a loss of derivatives with respect to the elliptic case depending directly on the structure of their singular spaces. We refer the reader to~\cite{karel,karel1} for a complete presentation of these subelliptic properties and examples of such subelliptic quadratic operators.

In the present work, we shall consider this class of hypoelliptic operators with zero singular space and study further the structure of the bottom of their spectra. More specifically, we shall see that  the first eigenvalue in the bottom of their spectra has always algebraic multiplicity one with an eigenspace spanned by a ground state of exponential type. We shall also give an explicit formula for the spectral gap which is computable via a simple algebraic calculation, and finally answer the question of long time behavior of the associated evolution equations by proving the property of exponential return to equilibrium for these quadratic systems.

Let us finally mention that the singular space theory  \cite{hps,hps1,hps2,karel,karel1} apply in various settings and that certain applications to the study of chains of oscillators and the generalized Langevin equation are discussed at the end of this article.

\section{Statements of the main results}
\label{sec:return to equil}

We begin by recalling miscellaneous facts and notations about quadratic operators. In all the following,
\begin{eqnarray*}
q : \rr_x^n \times \rr_{\xi}^n &\rightarrow& \cc\\
 (x,\xi) & \mapsto & q(x,\xi),
\end{eqnarray*}
stands for a complex-valued quadratic form with a non-negative real part
\begin{equation}\label{inf1}
\textrm{Re }q(x,\xi) \geq 0, \ (x,\xi) \in \rr^{2n}, n \in \nn^*.
\end{equation}
We know from \cite{mehler} (p.425-426) that the
maximal closed realization of  the operator $q^w(x,D_x)$, i.e., the operator on $L^2(\rr^n)$ with domain
$$D(q)=\big\{u \in L^2(\rr^n) :  q^w(x,D_x) u \in L^2(\rr^n)\big\},$$
coincides with the graph closure of its restriction to $\mathscr{S}(\rr^n)$,
$$q^w(x,D_x) : \mathscr{S}(\rr^n) \rightarrow \mathscr{S}(\rr^n)$$
and that this operator is maximally accretive
$$\textrm{Re}\big(q^w(x,D_x)u,u\big)_{L^2} \geq 0, \ u \in D(q).$$ 
Associated to the quadratic symbol $q$ is the numerical range $\Sigma(q)$ defined as the closure in the
complex plane of all its values,
\begin{equation}\label{9}
\Sigma(q)=\overline{q(\rr_x^n \times \rr_{\xi}^n)}. 
\end{equation}
The Hamilton map $F \in M_{2n}(\cc)$ associated to the quadratic form $q$
is the unique map defined by the identity 
\begin{equation}\label{10}
q\big{(}(x,\xi);(y,\eta) \big{)}=\sigma \big{(}(x,\xi),F(y,\eta) \big{)}, \ (x,\xi) \in \rr^{2n},  (y,\eta) \in \rr^{2n}; 
\end{equation}
where $q\big{(}\textrm{\textperiodcentered};\textrm{\textperiodcentered} \big{)}$ stands for the polarized form
associated to the quadratic form $q$ and $\sigma$ is the canonical
symplectic form on $\rr^{2n}$,
\begin{equation}\label{11}
\sigma \big{(}(x,\xi),(y,\eta) \big{)}=\xi.y-x.\eta, \ (x,\xi) \in \rr^{2n},  (y,\eta) \in \rr^{2n}.
\end{equation}
It directly follows from the definition of the Hamilton map $F$ that
its real part and its imaginary part
$$\textrm{Re } F=\frac{1}{2}(F+\overline{F}) \textrm{ and } \textrm{Im }F=\frac{1}{2i}(F-\overline{F}),$$ 
are the Hamilton maps associated
to the quadratic forms $\textrm{Re } q$ and $\textrm{Im }q$, respectively. One can notice from (\ref{10}) that a
Hamilton map is always skew-symmetric with respect to $\sigma$. This is just a consequence of the
properties of skew-symmetry of the symplectic form and symmetry of the polarized form
\begin{equation}\label{12}
\forall X,Y \in \rr^{2n}, \ \sigma(X,FY)=q(X;Y)=q(Y;X)=\sigma(Y,FX)=-\sigma(FX,Y).
\end{equation}
Associated to the quadratic symbol $q$ is the singular space $S$ defined in~\cite{hps} as the subvector space in the phase space equal to 
the following intersection of kernels
\begin{equation}\label{h1}
S=\Big(\bigcap_{j=0}^{+\infty}\textrm{Ker}\big[\textrm{Re }F(\textrm{Im }F)^j \big]\Big) \cap \rr^{2n}, 
\end{equation}
where the notations $\textrm{Re } F$ and $\textrm{Im }F$ stand respectively for the real and imaginary parts of the Hamilton map associated to $q$.
Notice that the Cayley-Hamilton theorem applied to $\textrm{Im }F$ shows 
$$(\textrm{Im }F)^k X \in \textrm{Vect}\big(X,...,(\textrm{Im }F)^{2n-1}X\big), \ X \in \rr^{2n}, \ k \in \nn,$$
where $\textrm{Vect}\big(X,...,(\textrm{Im }F)^{2n-1}X\big)$ is the vector space spanned by the vectors $X$,...,
$(\textrm{Im }F)^{2n-1}X$.
The singular space is therefore equal to the finite intersection of the kernels
\begin{equation}\label{h1bis}
S=\Big(\bigcap_{j=0}^{2n-1}\textrm{Ker}\big[\textrm{Re }F(\textrm{Im }F)^j \big]\Big) \cap \rr^{2n}. 
\end{equation}
As mentioned above, when the quadratic symbol $q$ satisfies a condition of partial ellipticity along its singular space $S$,
\begin{equation}\label{partial ellipticity}
(x,\xi) \in S, \ q(x,\xi)=0 \Rightarrow (x,\xi)=0,
\end{equation}
Theorem~1.2.2 in~\cite{hps} gives a complete description for the spectrum of the quadratic operator $q^w(x,D_x)$ which is only composed of eigenvalues with finite algebraic multiplicity 
\begin{equation}\label{sm61}
\sigma\big{(}q^w(x,D_x)\big{)}=\Big\{ \sum_{\substack{\lambda \in \sigma(F) \\  -i \lambda \in \cc_+
\cup (\Sigma(q|_S) \setminus \{0\})
} }
{\big{(}r_{\lambda}+2 k_{\lambda}
\big{)}(-i\lambda) : k_{\lambda} \in \nn}
\Big\}, 
\end{equation}
where $F$ is the Hamilton map associated to the quadratic form $q$, $r_{\lambda}$ is the dimension of the space of generalized eigenvectors of $F$ in $\cc^{2n}$
belonging to the eigenvalue $\lambda \in \cc$,
$$\Sigma(q|_S)=\overline{q(S)} \textrm{ and } \cc_+=\{z \in \cc : \textrm{Re }z>0\}.$$
In the present paper, we study evolution problems associated to accretive quadratic operators $q^w(x,D_x)$ with zero singular space $S=\{0\}.$ 
Notice that in this case $\overline{q(S)}=\{0\}$
and the condition of partial ellipticity along the singular space trivially holds. The spectrum is then reduces to 
\begin{equation}\label{sm6}
\sigma\big{(}q^w(x,D_x)\big{)}=\Big\{ \sum_{\substack{\lambda \in \sigma(F) \\  -i \lambda \in \cc_+
} }
{\big{(}r_{\lambda}+2 k_{\lambda}
\big{)}(-i\lambda) : k_{\lambda} \in \nn}
\Big\}.
\end{equation}
Define 
\begin{equation}\label{vp0}
\mu_0=\sum_{\substack{\lambda \in \sigma(F) \\  -i \lambda \in \cc_+
} }
 -i\lambda r_{\lambda}
\end{equation}
and 
\begin{equation}\label{rate}
\tau_0=\min_{\substack{\lambda \in \sigma(F) \\  -i \lambda \in \cc_+}} \textrm{Re}\big(2(-i\lambda)\big)
=2 \min_{\substack{\lambda \in \sigma(F) \\   \textrm{Im }\lambda>0}} \textrm{Im }\lambda,
\end{equation}
Our first result shows that the first eigenvalue in the bottom of the spectrum $\mu_0$ has always algebraic multiplicity one with an eigenspace spanned by a ground state of exponential type and that the spectral gap is exactly given by the positive rate $\tau_0>0$. 

\bigskip

\begin{theorem}\label{th1}
Let $q^w(x,D_x)$ be a quadratic operator whose symbol  
\begin{eqnarray*}
q : \rr_x^n \times \rr_{\xi}^n &\rightarrow& \cc\\
 (x,\xi) & \mapsto & q(x,\xi),
\end{eqnarray*}
is a complex-valued quadratic form with a non-negative real part $\emph{\textrm{Re }}q \geq 0$ and a zero singular space $S=\{0\}$. Then, the first eigenvalue in the bottom of the spectrum defined in \emph{(\ref{vp0})} has algebraic multiplicity one and the eigenspace
$$\emph{\textrm{Ker}}\big(q^w(x,D_x)-\mu_0\big)=\cc u_0,$$
is spanned by a ground state of exponential type 
$$u_0(x)=e^{-a(x)} \in \mathscr{S}(\rr^n),$$ 
where $a$ is a complex-valued quadratic form on $\rr^n$ whose real part is positive definite. Moreover, the spectral gap of the operator $q^w(x,D_x)$ is exactly given by the positive rate $\tau_0>0$ defined in \emph{(\ref{rate})},
$$\sigma\big(q^w(x,D_x)\big) \setminus \{\mu_0\} \subset \{z \in \cc : \emph{\textrm{Re }}z \geq \emph{\textrm{Re }}\mu_0 +\tau_0\}.$$
\end{theorem}

\bigskip

\begin{figure}[ht]
\caption{Notice that the first eigenvalue in the bottom of the spectrum is not necessarily real.}
\centerline{\includegraphics[scale=0.75]{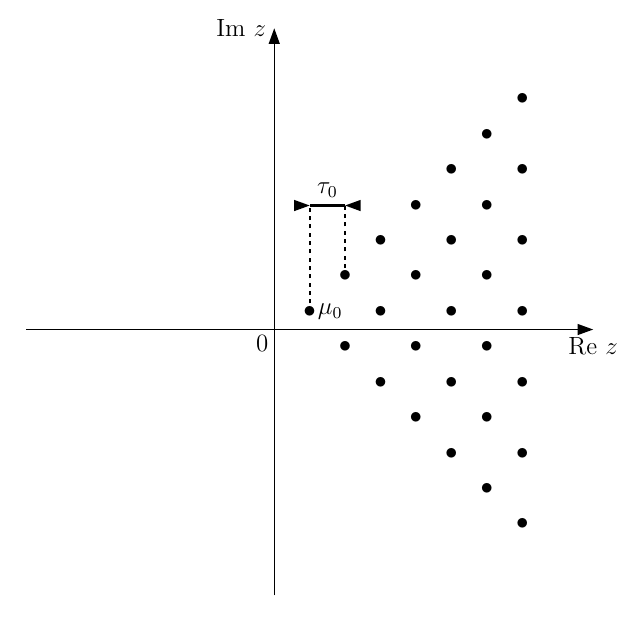}}
\end{figure}

This result of simplicity for the first eigenvalue is a result of Perron-Frobenius type. However, we notice that the ground state is non-vanishing but complex-valued since  in general the quadratic form $a$ is complex-valued. Additional assumptions are needed to ensure the positivity of the ground state (see Theorem~\ref{th3}).
Let us underline that operators satisfying the assumptions of Theorem~\ref{th1} are not elliptic in general and that their symbol may fail to satisfy the ellipticity condition
$$(x,\xi) \in \rr^{2n}, \ q(x,\xi)=0 \Rightarrow (x,\xi)=0.$$
However, the assumption about the singular space $S=\{0\}$ ensures that such an operator enjoys noticeable subelliptic (hypoelliptic) properties
\begin{equation}\label{lol1}
\big\|\big(\langle(x,\xi)\rangle^{2(1-\delta)}\big)^w u\big\|_{L^2} \leq C(\|q^w(x,D_x) u\|_{L^2}+\|u\|_{L^2}),
\end{equation}
with $\langle(x,\xi)\rangle=(1+|x|^2+|\xi|^2)^{1/2}$ and a loss of derivatives with respect to the elliptic case $0 \leq \delta<1$. More specifically, consider 
\begin{eqnarray*}
q : \rr_x^n \times \rr_{\xi}^n &\rightarrow& \cc\\
 (x,\xi) & \mapsto & q(x,\xi),
\end{eqnarray*}
a complex-valued quadratic form with a non-negative real part
$$\textrm{Re }q(x,\xi) \geq 0, \ (x,\xi) \in \rr^{2n}, \ n \in \nn^*,$$
and zero singular space $S=\{0\}$. According to the definition of the singular space (\ref{h1bis}), there exists a smallest integer $0 \leq k_0 \leq 2n-1$ such that
\begin{equation}\label{er133}
\Big(\bigcap_{j=0}^{k_0}\textrm{Ker}\big[\textrm{Re }F(\textrm{Im }F)^j \big]\Big) \cap \rr^{2n}=\{0\},
\end{equation}
where $F$ stands for the Hamilton map of $q$. Theorem~1.2.1 in~\cite{karel} shows that the quadratic operator fulfills a global subelliptic estimate with loss of $\delta=2k_0/(2k_0+1)$ derivatives with respect to the elliptic case, that is, there exists $C>0$ such that for all $u \in D(q)$, 
\begin{equation}\label{est1}
\big\|\big(\langle(x,\xi)\rangle^{2/(2k_0+1)}\big)^w u\big\|_{L^2} \leq C\big(\|q^w(x,D_x) u\|_{L^2}+\|u\|_{L^2}\big),
\end{equation}
with
$$D(q)=\{u \in L^2(\rr^n) : q^w(x,D_x)u \in L^2(\rr^n)\}.$$
Those subelliptic estimates are a key argument for getting sharp resolvent estimates and proving the property of exponential return to equilibrium. Next result proves this property of exponential return to equilibrium and provides an exact formula for the optimal rate of convergence:

\bigskip

\begin{theorem}\label{th2} 
Let $q^w(x,D_x)$ be a quadratic operator satisfying the assumptions of Theorem~\emph{\ref{th1}}. By using the notations introduced in \emph{(\ref{vp0})} and \emph{(\ref{rate})}, we consider the operator 
$$Q=q^w(x,D_x)-\mu_0.$$
Then, for all $0 \leq \tau <\tau_0$, there exists a positive constant $C>0$ such that 
$$\forall t \geq 0, \ \|e^{-tQ}-\Pi_0\|_{\mathcal{L}(L^2)} \leq Ce^{-\tau t},$$ 
where $\Pi_0$ is the rank-one spectral projection associated with the simple eigenvalue zero of the operator $Q$ and $\|\cdot\|_{\mathcal{L}(L^2)}$ stands for the norm of bounded operators on $L^2(\rr^n)$. 
\end{theorem}

\bigskip

Let $q^w(x,D_x)$ be a quadratic operator satisfying to the assumptions of Theorems~\ref{th1} or \ref{th2}, and $\mu_0$ be the first eigenvalue in the bottom of the spectrum. Theorem~\ref{th1} shows that
$$\mu_0=\sum_{\substack{\lambda \in \sigma(F) \\  -i \lambda \in \cc_+
} }
 -i\lambda r_{\lambda} \in \cc,$$
where $F$ stands for the Hamilton map of the quadratic form $q$. 
We shall now consider the specific case when this quadratic operator is real in the sense that $q^w(x,D_x)u$ is a real-valued function whenever $u$ is a real-valued function.
We begin by checking that this assumption ensures that $\mu_0$ is real and that the quadratic form $a$ defining the ground state 
$$u_0(x)=e^{-a(x)} \in \mathscr{S}(\rr^n), \quad q^w(x,D_x)u_0=\mu_0 u_0,$$ 
is then positive definite. Indeed, by using that the operator is real and passing to the complex conjugate in the equation
$$q^w(x,D_x)e^{-a(x)}=\mu_0 e^{-a(x)},$$
we obtain that
$$q^w(x,D_x)e^{-\overline{a}(x)}=\overline{\mu}_0 e^{-\overline{a}(x)}.$$
It shows that $\overline{\mu}_0$ is also an eigenvalue for the operator $q^w(x,D_x)$. Since by Theorem~\ref{th1}, we have
$$\sigma\big(q^w(x,D_x)\big) \setminus \{\mu_0\} \subset \{z \in \cc : \textrm{Re }z \geq \textrm{Re }\mu_0 +\tau_0\},$$
with $\tau_0>0$, $\textrm{Re }\mu_0=\textrm{Re }\overline{\mu}_0$ implies that the eigenvalue $\mu_0$ is real. Furthermore, notice that $e^{-a}$ and $e^{-\overline{a}}$ are therefore two eigenvectors associated to the eigenvalue $\mu_0$. We deduce from Theorem~\ref{th1} that there exists $\lambda_1, \lambda_2 \in \cc$, not both equal to zero such that for any $x \in \rr^n$,
$$\lambda_1 e^{-a(x)}+\lambda_2 e^{-\overline{a}(x)}=0.$$
Recalling that $a$ is a quadratic form, we obtain with $x=0$ that $\lambda_1=-\lambda_2$, which
implies successively that $e^{-a}-e^{-\overline{a}}=0$, $e^{-i b}-e^{i b}=0$, $e^{2ib}=1$, $\sin(2b)=0$, $\cos(2b)=1$,
if $b$ stands for the imaginary part of the quadratic form $a$. This shows that the quadratic form $b$ is constant 
$$0=\partial_{x_j}\big(\sin(2b)\big)=2\cos(2b)\partial_{x_j}b=2\partial_{x_j}b.$$
and therefore identically equal to zero. The quadratic form $a$ is therefore a positive definite quadratic form.

On the other hand, notice that the adjoint operator (see \cite{mehler}, p.426) is actually given by the quadratic operator
$$q^w(x,D_x)^*=\overline{q}^w(x,D_x),$$
whose symbol is the complex conjugate symbol of $q$. It is therefore a complex-valued quadratic form with non-negative real part and a zero singular space for which Theorem~\ref{th1} applies.

Then, $\mu_0$ is also the first eigenvalue in the bottom of the spectrum for the quadratic operator $q^w(x,D_x)^*$, and we know from Theorem~\ref{th1} that the eigenspaces associated with the eigenvalue $\mu_0$ for both operators $q^w(x,D_x)$ and $q^w(x,D_x)^*$ are one-dimensional subvector spaces with ground states of exponential type. We shall assume further that the two operators have same ground state
\begin{equation}\label{y11}
\textrm{Ker}\big(q^w(x,D_x)-\mu_0\big)=\textrm{Ker}\big(q^w(x,D_x)^*-\mu_0\big)=\cc u_0 \subset \mathscr{S}(\rr^n),
\end{equation} 
with $u_0(x)=e^{-a(x)}$, $x \in \rr^n$, where $a$ is a positive definite quadratic form on $\rr^n$. Under those assumptions, one can specify further the result given in Theorem~\ref{th2}:

\bigskip

\begin{theorem}\label{th3} 
Let $q^w(x,D_x)$ be a quadratic operator satisfying the assumptions of Theorem~\emph{\ref{th1}}. Assume that this quadratic operator is real and satisfies \emph{(\ref{y11})}.  Using the notations introduced in \emph{(\ref{vp0})} and \emph{(\ref{rate})}, we consider the operator 
$$Q=q^w(x,D_x)-\mu_0.$$
Then, for all $0 \leq \tau <\tau_0$, there exists a positive constant $C>0$ such that 
$$\forall t \geq 0, \forall u \in L^2(\rr^n), \ \|e^{-tQ}u-c_uu_0\|_{L^2(\rr^n)} \leq Ce^{-\tau t}\|u\|_{L^2(\rr^n)},$$ 
where $c_u$ is the $L^2(\rr^n)$ scalar product of $u$ and $u_0/\|u_0\|_{L^2(\rr^n)}^2$,
$$c_u=\|u_0\|_{L^2(\rr^n)}^{-2}(u,u_0)_{L^2(\rr^n)}.$$
\end{theorem}

\bigskip

We finally recall that the singular space theory \cite{hps,hps1,hps2,karel,karel1} was mainly inspired by the seminal work on the Fokker-Planck equation~\cite{HeSjSt} and aims at extending to wider classes of doubly characteristic operators some results proved for Fokker-Planck operators in~\cite{HeHiSj,HeHiSj2,HeHiSj3,HeSjSt}. After preliminary advances in this direction in~\cite{mz}, the main algebraic fact discovered in~\cite{hps} was to notice that the symbols of operators satisfying the assumptions of Theorem~\ref{th1} have real parts which may not be positive definite in general; however, when the real part of these quadratic symbols are averaged by the flow of the Hamilton vector field associated to their imaginary parts $H_{\textrm{Im}q}$,
\begin{equation}\label{hi1}
\langle \textrm{Re }q\rangle_T(X)=\frac{1}{2T}\int_{-T}^T{\textrm{Re }q(e^{tH_{\textrm{Im}q}}X)dt} \gg 0,
\end{equation}
with $X=(x,\xi) \in \rr^{2n}$, quadratic forms $\langle \textrm{Re }q\rangle_T$ become positive definite for any $T>0$. This averaging condition (\ref{hi1}) is a key property in the works~\cite{HeHiSj,HeHiSj2,HeHiSj3,HeSjSt}. One of the main interest of the singular space theory is to provide an easily computable algebraic condition on symbols of general quadratic operators for checking that the averaging property (\ref{hi1}) holds.  The remainder of this article is organized as follows. Next section is devoted to the proofs of Theorem~\ref{th1}, \ref{th2} and \ref{th3} whereas various applications of these results to the study of chains of oscillators and the generalized Langevin equation are discussed in Section~\ref{applications}.

\bigskip

\noindent
\textbf{Acknowledgement.} The third author would like to express his gratefulness to Johannes Sj\"ostrand for enriching remarks and comments on the present work.

\section{Proof of the main results}

\noindent
This section is devoted to the proofs of Theorems~\ref{th1}, \ref{th2} and \ref{th3}.

\subsection{Proof of Theorem~\ref{th1}} We begin by noticing that the set 
\begin{equation}\label{mi1}
\{\lambda \in \cc : \lambda \in \sigma(F), \ -i \lambda \in \cc_+\}=\{\lambda \in \cc : \lambda \in \sigma(F), \ \textrm{Im }\lambda>0\},
\end{equation}
is non-empty. Indeed, the work~\cite{hps} gives an alternative description of the singular space in terms of eigenspaces of the Hamilton map $F$
associated to its real eigenvalues. More specifically, when $q$ is a complex-valued quadratic form with a non-negative real part fulfilling the partial ellipticity condition on its singular space
$$(x,\xi) \in S, \ q(x,\xi)=0 \Rightarrow (x,\xi)=0,$$
which is obviously satisfied when $S=\{0\}$, then the set of real eigenvalues of the Hamilton map $F$ can be written as
$$\sigma(F) \cap \rr =\{\lambda_1,...,\lambda_r,-\lambda_1,...,-\lambda_r\},$$
with $\lambda_j \neq 0$ and $\lambda_j \neq \pm \lambda_k$  if $j \neq k$.
The singular space is then (\cite{hps}, p.807) the direct sum of the symplectically orthogonal spaces
\begin{equation}\label{sm5bis}
S=S_{\lambda_1} \oplus^{\sigma \perp} S_{\lambda_2} \oplus^{\sigma \perp}... \oplus^{\sigma \perp} S_{\lambda_r}, 
\end{equation}
where $S_{\lambda_{j}}$ is the symplectic space
\begin{equation}\label{sm5bis1}
S_{\lambda_j}=\big(\textrm{Ker}(F -\lambda_j) \oplus \textrm{Ker}(F+\lambda_j) \big) \cap \rr^{2n}. 
\end{equation}
Recalling from Proposition~4.4 in~\cite{mehler} that the subvector space 
$$\textrm{Ker}(F -\lambda_j) \oplus \textrm{Ker}(F+\lambda_j),$$ 
is the complexification of its intersection with $\rr^{2n}$, condition $S=\{0\}$ therefore implies that the Hamilton map $F$ has no real eigenvalue
$$\sigma(F) \cap \rr= \emptyset.$$ 
On the other hand, we know from~\cite{hormander} (Lemma~21.5.2) that $V_{\lambda}$ the space of generalized eigenvectors of $F$ belonging to the eigenvalue $\lambda \in \cc$ is the dual with respect to the symplectic form of the space of generalized eigenvectors $V_{-\lambda}$ belonging to the eigenvalue $-\lambda$. It follows that 
$$\lambda \in \sigma(F) \Leftrightarrow -\lambda \in \sigma(F).$$  
This implies that the set 
$$\{\lambda \in \cc : \lambda \in \sigma(F), \ -i \lambda \in \cc_+\}=\{\lambda \in \cc : \lambda \in \sigma(F), \ \textrm{Im }\lambda>0\},$$ 
is non-empty, and that the quantities $\mu_0 \in \cc$ and $\tau_0>0$ are well-defined in (\ref{vp0}) and (\ref{rate}). Since from (\ref{sm6}) the spectrum of the operator $q^w(x,D_x)$ is only composed of the following eigenvalues with finite algebraic multiplicity
\begin{equation}\label{yiw1}
\sigma\big{(}q^w(x,D_x)\big{)}=\Big\{ \sum_{\substack{\lambda \in \sigma(F) \\  -i \lambda \in \cc_+
} }
{\big{(}r_{\lambda}+2 k_{\lambda}
\big{)}(-i\lambda) : k_{\lambda} \in \nn}
\Big\},
\end{equation}
we notice that the first eigenvalue in the bottom of the spectrum is equal to 
$$\mu_0=\sum_{\substack{\lambda \in \sigma(F) \\  -i \lambda \in \cc_+} } -i\lambda r_{\lambda}$$
and we easily identify the spectral gap to be the positive rate
$$\tau_0=\min_{\substack{\lambda \in \sigma(F) \\  -i \lambda \in \cc_+}} \textrm{Re}\big(2(-i\lambda)\big)=2 \min_{\substack{\lambda \in \sigma(F) \\   \textrm{Im }\lambda>0}} \textrm{Im }\lambda>0.$$
It remains to check that this eigenvalue $\mu_0$ has algebraic multiplicity one with an eigenspace
$$\textrm{Ker}\big(q^w(x,D_x)-\mu_0\big)=\cc u_0,$$
spanned by a ground state of exponential type 
$$u_0(x)=e^{-a(x)} \in \mathscr{S}(\rr^n),$$ 
where $a$ is a complex-valued quadratic form on $\rr^n$ whose real part is positive definite.

In order to do so, we shall first consider the case when the real part of the quadratic symbol $\textrm{Re }q$ is positive definite. In this specific case, the symbol $q$ is elliptic on the whole phase space $\rr^{2n}$,
$$(x,\xi) \in \rr^{2n}, \ q(x,\xi)=0 \Rightarrow (x,\xi)=0$$
and one can refer to the original work by J.~Sj\"ostrand on elliptic quadratic differential operators~\cite{sjostrand}. Following this fundamental work on which relies the analysis led in~\cite{hps}, we consider the positive Lagrangian plane (Proposition~3.3 in~\cite{sjostrand}),
$$V^+=\bigoplus_{\substack{\lambda \in \sigma(F) \\  -i \lambda \in \cc_+}}V_{\lambda},$$
where $V_{\lambda}$ is the space of generalized eigenvectors belonging to the eigenvalue $\lambda$ of the Hamilton map $F$ associated to the quadratic symbol $q$. We recall that a complex subspace $\Lambda \subset \cc^{2n}$ of dimension $n$ is called Lagrangian if the canonical symplectic form $\sigma$ vanishes identically on $\Lambda$. Furthermore, a Lagrangian plane $\Lambda \subset \cc^{2n}$ is said to be positive if 
$$-i\sigma(X,\overline{X})>0,$$ 
for all $0 \neq X=(x,\xi) \in \Lambda.$
According to (3.5) in~\cite{sjostrand} (p.95), there exists a symmetric matrix $B^+$ whose imaginary part $\textrm{Im }B^+$ is positive definite such that 
$$V^+=\{(x,B^+x) : x \in \cc^n\}.$$
Define the two quadratic forms
$$b^+(x)=\langle x,B^+x \rangle, \ x \in \rr^n; \ a(x)=-\frac{1}{2}ib^+(x), \ x \in \rr^n.$$
Notice that the real part of the quadratic form $a$ is positive definite. Keeping on following the analysis led in~\cite{sjostrand} (p.98), on can then use a Jordan decomposition for the Hamilton map $F$ in order to find a canonical transformation $\mathcal{K}$ of $\cc^{2n}$ such that
$$\tilde{q}=2\sum_{\substack{\lambda \in \sigma(F) \\  -i \lambda \in \cc_+} }\Big(\sum_{j=1}^{r_{\lambda}}\lambda x_{j,\lambda}\xi_{j,\lambda}+\sum_{j=1}^{r_{\lambda}-1}\gamma_{j,\lambda}x_{j,\lambda}\xi_{j+1,\lambda}\Big),$$
with $\tilde{q}=q \circ \mathcal{K}^{-1}$. Here, $x_{j,\lambda},\xi_{j,\lambda}$, with $1 \leq j \leq r_{\lambda}$, $\lambda \in \sigma(F)$ and $-i \lambda \in \cc_+$ stands for a relabeling of the standard real symplectic coordinates $(x,\xi) \in \rr^{2n}$ and $\gamma_{j,\lambda} \in \{0,1\}$ with $\gamma_{0,\lambda}=\gamma_{r_{\lambda},\lambda}=0$. We recall that $r_{\lambda}$ stands for the dimension of $V_{\lambda}$ the space of generalized eigenvectors associated to the eigenvalue $\lambda \in \cc$. This reduction is proved in~\cite{sjostrand} (p.98) in (3.15) and (3.16). Consider the two quadratic operators
\begin{equation}\label{nilpo1}
\tilde{q}_1^w(x,D_x)=\sum_{\substack{\lambda \in \sigma(F) \\  -i \lambda \in \cc_+} }\sum_{j=1}^{r_{\lambda}}\lambda \big( x_{j,\lambda}D_{x_{j,\lambda}}+ D_{x_{j,\lambda}}x_{j,\lambda}\big)
\end{equation} 
and
\begin{equation}\label{nilpo}
\tilde{q}_2^w(x,D_x)=2\sum_{\substack{\lambda \in \sigma(F) \\  -i \lambda \in \cc_+} }\sum_{j=1}^{r_{\lambda}-1}\gamma_{j,\lambda}x_{j,\lambda}D_{x_{j+1,\lambda}}.
\end{equation} 
Notice that these two operators commute and that
$$\tilde{q}^w(x,D_x)=\tilde{q}_1^w(x,D_x)+\tilde{q}_2^w(x,D_x).$$
Proposition~3.10 in~\cite{sjostrand} shows that every monomial is a generalized eigenvector of the operator
$$\tilde{q}^w(x,D_x) : P(\rr^n) \rightarrow P(\rr^n),$$
where $P(\rr^n)$ stands for the space of complex polynomials regarded as functions on $\rr^n$. More specifically, the spectrum of the operator $\tilde{q}^w(x,D_x)$ acting on the space $P(\rr^n)$ is exactly given by
\begin{equation}\label{sach1}
\Big\{ \sum_{\substack{\lambda \in \sigma(F) \\  -i \lambda \in \cc_+
} }
{\big{(}r_{\lambda}+2 k_{\lambda}
\big{)}(-i\lambda) : k_{\lambda} \in \nn}
\Big\}.
\end{equation} 
Proposition~3.10 in~\cite{sjostrand} also shows that the space of generalized eigenvectors 
$$E\big(\tilde{q}^w(x,D_x),P(\rr^n),\mu\big),$$ 
associated to an eigenvalue $\mu$ is finite-dimensional
$$\textrm{dim }E\big(\tilde{q}^w(x,D_x),P(\rr^n),\mu\big) <+\infty.$$
Following the proof of Proposition~3.10 in~\cite{sjostrand}, a direct computation shows that 
$$\sum_{j=1}^{r_{\lambda}}\lambda \big( x_{j,\lambda}D_{x_{j,\lambda}}+ D_{x_{j,\lambda}}x_{j,\lambda}\big)p_{\lambda}=-i\lambda\big(r_{\lambda}+2 k_{\lambda}\big)p_{\lambda},$$
for any homogeneous polynomial of degree $k_{\lambda}$ in the variables $x_{1,\lambda}$,..., $x_{r(\lambda),\lambda}$.
This implies that the monomials 
$$p(x)=\prod_{\substack{\lambda \in \sigma(F) \\  -i \lambda \in \cc_+
} }p_{\lambda}(x),$$
constitute a basis of eigenvectors for the operator $\tilde{q}_1^w(x,D_x)$ acting on $P(\rr^n)$,
$$\tilde{q}_1^w(x,D_x)p=\sum_{\substack{\lambda \in \sigma(F) \\  -i \lambda \in \cc_+}}\big(r_{\lambda}+2 k_{\lambda}\big)(-i\lambda)p.$$
According to this description, we notice that the first eigenvalue in the bottom of the spectrum for the operator $\tilde{q}_1^w(x,D_x)$ acting on the space $P(\rr^n)$ is the value $\mu_0$ defined in (\ref{vp0}). Furthermore, this eigenvalue $\mu_0$ has algebraic multiplicity 1. Indeed, all the non-negative integers $k_{\lambda}$ must necessarily be zero and the only possible eigenvector associated to $\mu_0$ is up to a non-zero constant the constant polynomial $1$. 
By using the expression (\ref{nilpo}), a direct computation shows that the operator $\tilde{q}_2^w(x,D_x)$ is nilpotent on any space of generalized eigenvectors 
$$E\big(\tilde{q}_1^w(x,D_x),P(\rr^n),\mu\big), \quad \mu \in \cc,$$
for the operator $\tilde{q}_1^w(x,D_x)$ seen as acting on the space $P(\rr^n)$, since the operator $\tilde{q}_2^w(x,D_x)$ is obviously nilpotent on any subvector space spanned by the monomials
$$\prod_{\substack{\lambda \in \sigma(F) \\  -i \lambda \in \cc_+
} }p_{\lambda}(x),$$
where $p_{\lambda}$ are homogeneous polynomials of degree $k_{\lambda}$ in the variables $x_{1,\lambda}$,..., $x_{r(\lambda),\lambda}$ satisfying
$$\mu=\sum_{\substack{\lambda \in \sigma(F) \\  -i \lambda \in \cc_+}}\big(r_{\lambda}+2 k_{\lambda}\big)(-i\lambda).$$
This proves in particular that $\mu_0$ is an eigenvalue with algebraic multiplicity 1 for the operator 
$$\tilde{q}^w(x,D_x) : P(\rr^n) \rightarrow P(\rr^n),$$
with eigenvector equal to the constant polynomial 1. According to \cite{sjostrand} (p.100), there exists a bijection
$$K : P(\rr^n)e^{ib^+(x)/2} \rightarrow P(\rr^n),$$
such that 
\begin{equation}\label{sach2}
\tilde{q}^w(x,D_x) \circ K =K \circ q^w(x,D_x).
\end{equation}
More specifically, according to (3.22), (3.23), (3.24) and (3.25) in~\cite{sjostrand}, this mapping may be written as
\begin{equation}\label{sach13}
K=K_3K_2K_1 : P(\rr^n)e^{ib^+(x)/2} \rightarrow P(\rr^n),
\end{equation} 
with some operators
\begin{equation}\label{sach3}
K_1 : P(\rr^n)e^{ib^+(x)/2} \rightarrow P(\rr^n), \quad  K_1u=e^{-ib^+(x)/2}u,
\end{equation}
\begin{equation}\label{sach4}
K_2 : P(\rr^n) \rightarrow P(\rr^n), \quad  K_2u=\mathcal{F}^{-1}\big(e^{ib^-(\xi)/2}\mathcal{F}u\big),
\end{equation}
and
\begin{equation}\label{sach5}
K_3 : P(\rr^n) \rightarrow P(\rr^n), \quad  (K_3u)(x)=u(Cx),
\end{equation}
where $\mathcal{F}$ is the Fourier transformation, $b^-$ a quadratic form and $C$ an invertible complex matrix.
Proposition~3.11 in~\cite{sjostrand} then establishes that the spectrum of the operator 
$$q^w(x,D_x) : P(\rr^n)e^{ib^+(x)/2} \rightarrow P(\rr^n)e^{ib^+(x)/2},$$
is also exactly given by (\ref{sach1}) and that for every eigenvalue the corresponding space of generalized eigenfunctions is finite dimensional. More specifically, we deduce from (\ref{sach2}), (\ref{sach13}), (\ref{sach3}), (\ref{sach4}) and (\ref{sach5}) that $\mu_0$ is an eigenvalue with algebraic multiplicity 1 for the operator 
$$q^w(x,D_x) : P(\rr^n)e^{ib^+(x)/2} \rightarrow P(\rr^n)e^{ib^+(x)/2},$$
with associated eigenvector $K^{-1}(1)$. An explicit computation shows that
$$K^{-1}(1)=e^{ib^+(x)/2}=e^{-a(x)},$$ 
with $a=-ib^+/2$. As a final step in the proof of Theorem~3.5, it is proved in~\cite{sjostrand} (p.102) that for any $\mu \in \cc$, the space of generalized eigenvectors associated to the eigenvalue $\mu$ for the operator
$$q^w(x,D_x) : P(\rr^n)e^{ib^+(x)/2} \rightarrow P(\rr^n)e^{ib^+(x)/2},$$
exactly coincides with the space of generalized eigenvectors associated to the eigenvalue $\mu$ for the operator
$$q^w(x,D_x) : L^2(\rr^n) \rightarrow L^2(\rr^n),$$
that is
$$E\big(q^w(x,D_x), P(\rr^n)e^{ib^+/2},\mu\big)=E\big(q^w(x,D_x), L^2(\rr^n),\mu\big).$$
This implies that when the real part of the quadratic symbol $\textrm{Re }q$ is positive definite, $\mu_0$ the first eigenvalue in the bottom of the spectrum of the operator $q^w(x,D_x)$ acting on $L^2(\rr^n)$ has algebraic multiplicity 1 with a ground state of exponential type
$$q^w(x,D_x)u_0=\mu_0 u_0,$$ 
with $u_0(x)=e^{-a(x)} \in \mathscr{S}(\rr^n),$ where $a$ is a complex-valued quadratic form on $\rr^n$, whose real part is positive definite. This proves Theorem~\ref{th1} in the special case when  the real part of the quadratic symbol $\textrm{Re }q$ is positive definite.

Let us now consider the general case when the quadratic symbol 
\begin{eqnarray*}
q : \rr_x^n \times \rr_{\xi}^n &\rightarrow& \cc\\
 (x,\xi) & \mapsto & q(x,\xi),
\end{eqnarray*}
is a complex-valued quadratic form with a non-negative real part $\textrm{Re }q \geq 0$ and a zero singular space $S=\{0\}$. We shall now inspect the analysis led in~\cite{hps,hps2} in order to prove Theorem~\ref{th1} in the general case. Notice that in this general case the real part of the quadratic symbol $\textrm{Re }q$ is not necessarily positive definite. However, as noticed in~\cite{hps} (p.809-810 and Proposition~2.0.1) and mentioned above, the fact that the singular space $S$ is reduced to zero implies that the average of the real part of the quadratic form $q$ by the flow generated by the Hamilton vector field of its imaginary part
$H_{\textrm{Im}q}$,
$$\langle \textrm{Re }q\rangle_T(X)=\frac{1}{2T}\int_{-T}^T{\textrm{Re }q(e^{tH_{\textrm{Im}q}}X)dt}, \ X=(x,\xi) \in \rr^{2n}$$
is positive definite for any $T>0$. We shall now inspect more specifically the proof of Proposition~3.1.1 in~\cite{hps}. Notice that we consider in~\cite{hps} complex-valued quadratic symbols with non-positive real parts. One can therefore apply all this analysis when considering the symbol $-q$ instead of $q$ and this accounts for sign differences when referring to results proved in~\cite{hps}. Following \cite{hps} (p.825-831), we shall discuss certain IR-deformations of the real phase space $\rr^{2n}$ where the averaging procedure along the flow defined by the Hamilton vector field of $\textrm{Im }q$ plays an important r\^ole. To that end, we work with the metaplectic FBI-Bargmann
transform
\begin{equation}\label{eq4aaa}
T u(x) = C \int_{\rr^n} e^{i\varphi(x,y)} u(y)\,dy,\quad x\in \cc^n, \ C>0,
\end{equation}
where we may choose
$$\varphi(x,y)=\frac{i}{2} (x-y)^2,$$
as in the standard Bargmann transform.
Other quadratic phase functions $\varphi$ such that $\textrm{Im } \varphi''_{y y}>0$ and $\det \varphi''_{x y}\neq 0$ are also
possible \cite{Sj95}.
It is well known that for a suitable choice of $C>0$, $T$ defines a unitary transformation
$$
T: L^2(\rr^n)\rightarrow H_{\Phi_0}(\cc^n),
$$
where
\begin{equation}\label{eq4.1aaa}
H_{\Phi_0}(\cc^n)={\rm Hol}(\cc^n)\cap L^2\big(\cc^n,
e^{-2\Phi_0(x)}L(dx)\big), 
\end{equation}
with
$$
\Phi_0(x)=\sup_{y \in \rr^n} -\textrm{Im } \varphi(x,y)=\frac{1}{2} \left(\textrm{Im }x\right)^2,
$$
and $L(dx)$ being the Lebesgue measure in $\cc^n$. Next we recall from~\cite{Sj95} that
\begin{equation}\label{eq5aaa}
Tq^w(x,D_x)u=Q_0Tu,\quad u\in \mathscr{S}(\rr^n),
\end{equation}
where $Q_0$ is a quadratic differential operator on $\cc^n$ whose
Weyl symbol $q_0$ satisfies
\begin{equation}\label{inv7aaa}
q_0\circ \kappa_T = q. 
\end{equation}
Here
\begin{equation}\label{eq5.1aaa}
\kappa_T: \cc^{2n}\ni \big(y,-\varphi'_y(x,y)\big) \mapsto \big(x,\varphi'_x(x,y)\big)\in \cc^{2n}, 
\end{equation}
is the complex linear canonical transformation associated to
$T$. Following~\cite{Sj95}, we recall that if we define
\begin{equation}\label{inv8aaa}
\Lambda_{\Phi_0}=\Big\{\Big(x,\frac{2}{i}\frac{\partial \Phi_0}{\partial
x}(x)\Big): x\in \cc^n \Big\}, 
\end{equation}
then we have
\begin{equation}\label{inv9aaa}
\Lambda_{\Phi_0}=\kappa_T(\rr^{2n}).
\end{equation}
When
$$\sigma = \sum_{j=1}^n d\xi_j \wedge dx_j,$$
is the complex symplectic (2,0)-form on $\cc^{2n}=\cc^n_x\times
\cc^n_{\xi}$ then the restriction $\sigma_{\Lambda_{\Phi_0}}$ of
$\sigma$ to $\Lambda_{\Phi_0}$ is real and nondegenerate. The map
$\kappa_T$ in (\ref{eq5.1aaa}) can therefore be viewed as a canonical
transformation between the real symplectic spaces $\rr^{2n}$ and
$\Lambda_{\Phi_0}$. Continuing to follow~\cite{Sj95}, we recall that when realizing $Q_0$ as an unbounded operator on
$H_{\Phi_0}(\cc^n)$, we may use first the contour integral
representation
$$
Q_0 u(x) = \frac{1}{(2\pi)^n}\int_{\theta=\frac{2}{i}\frac{\partial \Phi_0}{\partial
x}\left(\frac{x+y}{2}\right)} e^{i(x-y)\cdot \theta}
q_0\Big(\frac{x+y}{2},\theta\Big)u(y)\,dy\,d\theta,
$$
and then using that the symbol $q_0$ is holomorphic, by a contour
deformation we obtain the following formula for $Q_0$ as an unbounded operator on $H_{\Phi_0}(\cc^n)$,
\begin{equation}\label{eq5.5aaa}
Q_0 u(x) = \frac{1}{(2\pi)^n}\int_{\theta=\frac{2}{i}\frac{\partial \Phi_0}{\partial
x}\left(\frac{x+y}{2}\right)+it\overline{(x-y)}} e^{i(x-y)\cdot \theta}
q_0\Big(\frac{x+y}{2},\theta\Big)u(y)\,dy\,d\theta, 
\end{equation}
for any $t>0$. The two operators $Q_0: H_{\Phi_{0}}(\cc^n) \rightarrow H_{\Phi_{0}}(\cc^n)$ and
$$q^w(x,D_x) : L^2(\rr^n) \rightarrow L^2(\rr^n)$$
are unitarily equivalent. Thus, they both have the same discrete spectrum since the operator $q^w(x,D_x)$ has compact resolvent according to (\ref{est1}).  
Let $G=G_T$ be a real-valued quadratic form on $\rr^{2n}$ satisfying
\begin{equation}\label{eq6}
H_{\textrm{Im} q} G = -\textrm{Re }q+\langle{\textrm{Re  }q}\rangle_T.
\end{equation}
As in~\cite{HeHiSj}, we solve (\ref{eq6}) by setting
\begin{equation}\label{eq7}
G(X)=\int_{\rr} k_T(t) \textrm{Re }q(e^{tH_{\textrm{Im} q}}X)dt,
\end{equation}
where $k_T(t)=k(t/2T)$ and $k\in C(\rr \setminus \{0\})$ is the odd function given by
$$
k(t)=0 \textrm{ for } |t| \geq \frac{1}{2} \textrm{ and } k'(t)=-1 \textrm{ for } 0<|t| < \frac{1}{2}.
$$
Let us notice that $k$ and $k_T$ have a jump of size $1$ at the origin. Associated with $G$ is a linear IR-manifold defined for $0\leq \eps\leq \eps_0$ with $0<\eps_0 \ll 1$,
\begin{equation}\label{eq8}
\Lambda_{\eps G}=e^{i\eps H_G}(\rr^{2n}) \subset \cc^{2n},
\end{equation}
where $e^{i\eps H_G}$ stands for the flow generated by the linear Hamilton vector field $i\eps H_G$ taken at time 1.
It is well-known (see for instance sections 3 and 5 in~\cite{HeSjSt}) that
\begin{equation}\label{eq9aaa}
\kappa_T(\Lambda_{\eps
G})=\Lambda_{\widetilde{\Phi}_{\eps}}:=\Big\{\Big(x,\frac{2}{i}\frac{\partial
\widetilde{\Phi}_{\eps}}{\partial x}(x)\Big) : x\in \cc^n\Big\},
\end{equation}
where $\widetilde{\Phi}_{\eps}$ is a strictly plurisubharmonic quadratic form on $\cc^n$ satisfying
\begin{equation}\label{lin1}
\widetilde{\Phi}_{\eps}(x)=\Phi_0(x)+\eps G(\textrm{Re }x, -\textrm{Im }x)+\mathcal{O}(\eps^2\abs{x}^2).
\end{equation}
Associated with the function $\widetilde{\Phi}_{\eps}$ is the weighted space of
holomorphic functions $H_{\widetilde{\Phi}_{\eps}}(\cc^n)$
defined as in
(\ref{eq4.1aaa}). The operator $Q_0$ can also be defined as an unbounded operator
$$
Q_0: H_{\widetilde{\Phi}_{\eps}}(\cc^n) \rightarrow H_{\widetilde{\Phi}_{\eps}}(\cc^n),
$$
if we make a new contour deformation in (\ref{eq5.5aaa}) and set
\begin{equation}\label{eq10}
Q_0 u(x) = \frac{1}{(2\pi)^n}\int_{\theta=\frac{2}{i}\frac{\partial \widetilde{\Phi}_{\eps}}{\partial
x}\left(\frac{x+y}{2}\right)+it\overline{(x-y)}} e^{i(x-y)\cdot \theta}
q_0\Big(\frac{x+y}{2},\theta\Big)u(y)\,dy\,d\theta,
\end{equation}
for any $t>0$. By coming back to the real side by the FBI-Bargmann transform, the operator $Q_0$ acting on $H_{\widetilde{\Phi}_{\eps}}(\cc^n)$ can be viewed as the
unbounded operator on $L^2(\rr^n)$ with Weyl symbol
\begin{equation}\label{esk1}
q_{\eps}(X)=q(e^{i \eps H_G}X). 
\end{equation}
This symbol is quadratic and its real part is easily seen to be equal to
$$\textrm{Re }q_{\eps}(X)=\textrm{Re }q(X) +\eps H_{\textrm{Im}q} G(X)+\mathcal{O}(\eps^2\abs{X}^2).$$
We deduce from (\ref{eq6}) and the facts that the quadratic form $\langle \textrm{Re }q \rangle_T$ is positive definite and $\textrm{Re }q \geq 0$ that
\begin{equation}\label{eq10.1}
\textrm{Re } q_{\eps}(X)=(1-\eps)\textrm{Re }q(X) +\eps \langle \textrm{Re }q \rangle_T(X)+\mathcal{O}(\eps^2\abs{X}^2) \geq \frac{\eps}{C}\abs{X}^2, 
\end{equation}
for $C>1$ and $0<\eps \ll 1$.
We proved in~\cite{hps} (Lemma~3.1.1) that for any $0<\eps \leq \eps_0$, with $\eps_0>0$ sufficiently small, the spectrum of the quadratic operator $q_{\eps}^w(x,D_x)$ is actually independent of the parameter $0<\eps \leq \eps_0$ and is only composed by eigenvalues with finite algebraic multiplicity
$$\forall \ 0 <\eps \leq \eps_0,\ \sigma\big{(}q_{\eps}^w(x,D_x)\big{)}=\Big\{ \sum_{\substack{\lambda \in \sigma(F) \\  -i \lambda \in \cc_+
} }
{\big{(}r_{\lambda}+2 k_{\lambda}
\big{)}(-i\lambda) : k_{\lambda} \in \nn}
\Big\},$$ 
where $F$ is the Hamilton map associated to the quadratic form $q$, $r_{\lambda}$ is the dimension of the space of generalized eigenvectors of $F$ in $\cc^{2n}$
belonging to the eigenvalue $\lambda \in \cc$. The independence of the spectrum of the operator $q_{\eps}^w(x,D_x)$ with respect to the parameter $0<\eps \leq \eps_0$ is linked to the fact that the Hamilton map $F_{\eps}$ of the quadratic form $q_{\eps}$ is isospectral with the Hamilton map $F$ of the quadratic form $q$, 
$$F_{\eps}=e^{-i \eps H_G}Fe^{i \eps H_G},$$
since the symbols $q$ and $q_{\eps}$ are related by a canonical transformation. As a final step in~\cite{hps,hps2}, we proved that if the positive constant $\eps_0>0$ is chosen sufficiently small then the spectrum of the operator $q^w(x,D_x)$ is not only equal to the spectrum of the operator $q_{\eps}^w(x,D_x)$ for any $0<\eps \leq \eps_0$, but also that all the eigenvectors and generalized eigenvectors of the two operators $q^w(x,D_x)$ and $q_{\eps}^w(x,D_x)$ for any $0<\eps \leq \eps_0$ agree. These facts were proven in \cite{hps} (p.830-831) and \cite{hps2} (Proposition~2.1). Indeed, we recall from \cite{hps}  that the general theory shows that the operator 
\begin{equation}\label{invb1}
e^{-tQ_0}: H_{\Phi_0}(\cc^n) \rightarrow H_{\Phi_t}(\cc^n), \quad 0 < t\leq t_0, \ 0<t_0 \ll 1, 
\end{equation}
is bounded. Here, $\Phi_t(x)=\Phi(t,x)$ is a strictly plurisubharmonic quadratic form on $\cc^n$ depending smoothly on $t$ and solving the eikonal equation
\begin{equation}
\label{eikonal}
\left\lbrace
\begin{array}{c}
\displaystyle \frac{\partial \Phi}{\partial t}(t,x)+ \textrm{Re}\Big[q_0\Big(x,\frac{2}{i}\frac{\partial \Phi}{\partial x}(t,x)\Big)\Big]=0 \\
\Phi(t,\textrm{\textperiodcentered})|_{t=0}=\Phi_0.
\end{array} \right.
\end{equation}
Lemma~3.1.2 in \cite{hps} shows that
\begin{equation}\label{eq20.1aaa}
\forall \ 0<t \ll 1, \exists \alpha(t)>0, \  \Phi_{t}(x)\leq \Phi_0(x) - \alpha(t) \abs{x}^2,\quad x\in \cc^n. 
\end{equation}
We consider the restriction of the heat semigroup $e^{-tQ_0}$ to a generalized eigenspace $E_{\lambda_0,\Phi_0}\subset H_{\Phi_0}(\cc^{n})$ associated to an eigenvalue
$\lambda_0$ for the operator 
$$Q_0: H_{\Phi_{0}}(\cc^n) \rightarrow H_{\Phi_{0}}(\cc^n).$$
The space $E_{\lambda_0,\Phi_0}$ is finite-dimensional and the restriction of $Q_0-\lambda_0$ to $E_{\lambda_0,\Phi_0}$ is nilpotent.
Notice that the map $e^{-tQ_0}: E_{\lambda_0,\Phi_0} \rightarrow E_{\lambda_0,\Phi_0}$ is bijective for any $t\geq 0$.
Indeed, the generalized eigenspace $E_{\lambda_0,\Phi_0}$ is stable under the action of the operator $Q_0$ and its restriction to this finite-dimensional space
$$Q_0|_{E_{\lambda_0,\Phi_0}} : E_{\lambda_0,\Phi_0} \rightarrow E_{\lambda_0,\Phi_0},$$
is a bounded operator. This implies that the restriction of the semigroup to the space $(e^{-tQ_0})|_{E_{\lambda_0,\Phi_0}}$ coincides with
the exponential of the bounded operator $-tQ_0|_{E_{\lambda_0,\Phi_0}}$ which is always bijective.
It follows from (\ref{invb1}) and (\ref{eq20.1aaa}) that there exists $\eta>0$ such that any generalized eigenspace $E_{\lambda_0,\Phi_0}\subset H_{\Phi_0}(\cc^{n})$ for the operator 
$Q_0$ acting on $H_{\Phi_{0}}(\cc^n)$ satisfies
$$E_{\lambda_0,\Phi_0} \subset  H_{\Phi_{t_0}}(\cc^n) \subset H_{\Phi_0-\eta \abs{x}^2}(\cc^{n}).$$
In particular, we obtain from (\ref{lin1}) that any generalized eigenspace $E_{\lambda_0,\Phi_0}$ is included in the space $H_{\tilde{\Phi}_{\eps}}(\cc^n)$ when $0<\eps \leq \eps_0$ with $0<\eps_0 \ll 1$. It follows that any function in $E_{\lambda_0,\Phi_0}$ is a generalized eigenvector for the operator 
$$Q_0: H_{\widetilde{\Phi}_{\eps}}(\cc^n) \rightarrow H_{\widetilde{\Phi}_{\eps}}(\cc^n),$$
when $0<\eps \leq \eps_0$. Conversely, we have
\begin{equation}\label{invb1bbb}
e^{-tQ_0}: H_{\widetilde{\Phi}_{\eps}}(\cc^n) \rightarrow H_{\widetilde{\Phi}_{\eps,t}}(\cc^n), \quad 0 < t\leq t_0,
\end{equation}
with $0<t_0 \ll 1$, where $\widetilde{\Phi}_{\eps,t}$ is a quadratic form on $\cc^n$ depending smoothly on $t\geq 0$ and $0 \leq \eps \ll 1$ satisfying to the eikonal equation (\ref{eikonal}) along with the initial condition
$$\widetilde{\Phi}_{\eps,t}(x)|_{t=0}=\widetilde{\Phi}_{\eps}(x).$$
It follows from (\ref{lin1}) and (\ref{eikonal}) that
$$\widetilde{\Phi}_{\eps,t}(x)=\Phi_t(x)+\mathcal{O}(\eps \abs{x}^2),$$
where the implicit constant is uniform in $0\leq t\leq t_0$ when $0< t_0 \ll 1$.
Let $E_{\lambda_0,\widetilde{\Phi}_{\eps}}\subset H_{\widetilde{\Phi}_{\eps}}(\cc^{n})$ be the generalized eigenspace associated to an eigenvalue $\lambda_0$ for the operator 
$$Q_0: H_{\widetilde{\Phi}_{\eps}}(\cc^n) \rightarrow H_{\widetilde{\Phi}_{\eps}}(\cc^n).$$
The very same arguments as before show that there exists $\eta>0$ such that any generalized eigenspace $E_{\lambda_0,\widetilde{\Phi}_{\eps}}\subset H_{\widetilde{\Phi}_{\eps}}(\cc^{n})$ for the operator $Q_0$ acting on $H_{\widetilde{\Phi}_{\eps}}(\cc^n)$ satisfies
$$E_{\lambda_0,\widetilde{\Phi}_{\eps}} \subset  H_{\widetilde{\Phi}_{\eps,t_0}}(\cc^n) \subset H_{\Phi_{t_0}+\mathcal{O}(\eps \abs{x}^2)} \subset H_{\Phi_0-\eta \abs{x}^2+\mathcal{O}(\eps \abs{x}^2)}(\cc^{n})\subset  H_{\Phi_{0}}(\cc^n),$$
when $0<\eps \leq \eps_0$ with $0<\eps_0 \ll 1$. It follows that any function in $E_{\lambda_0,\widetilde{\Phi}_{\eps}}$ is also a generalized eigenvector for the operator 
$Q_0: H_{\Phi_0}(\cc^n) \rightarrow H_{\Phi_{0}}(\cc^n)$ when $0<\eps \leq \eps_0$.
This shows that the spectra and the generalized eigenvectors for the operator $Q_0$ acting on $H_{\Phi_0}(\cc^{n})$ or  $H_{\tilde{\Phi}_{\eps}}(\cc^n)$ with $0<\eps \leq \eps_0$ agree. By coming back to the real side, this property holds also true for the operator $q^w(x,D_x)$ and $q_{\eps}^w(x,D_x)$ when  $0<\eps \leq \eps_0$.
This implies that one can deduce the result of Theorem~\ref{th1} in the general case when the quadratic symbol has a non-negative real part $\textrm{Re }q \geq 0$ and a zero 
 singular space $S=\{0\}$ from the one proved previously in the particular case when the real part of the symbol is positive definite. This ends the proof of Theorem~\ref{th1}.

\bigskip

\subsection{Proof of Theorem~\ref{th2}}
We consider 
\begin{eqnarray*}
q : \rr_x^n \times \rr_{\xi}^n &\rightarrow& \cc\\
 (x,\xi) & \mapsto & q(x,\xi),
\end{eqnarray*}
a complex-valued quadratic form with a non-negative real part
$$\textrm{Re }q(x,\xi) \geq 0, \ (x,\xi) \in \rr^{2n}, \ n \in \nn^*,$$
and zero singular space $S=\{0\}$. According to the definition of the singular space (\ref{h1bis}), there exists a smallest integer $0 \leq k_0 \leq 2n-1$ such that
\begin{equation}\label{er1}
\Big(\bigcap_{j=0}^{k_0}\textrm{Ker}\big[\textrm{Re }F(\textrm{Im }F)^j \big]\Big) \cap \rr^{2n}=\{0\},
\end{equation}
where $F$ stands for the Hamilton map of $q$ and Theorem~1.2.1 in~\cite{karel} shows that the operator $q^w(x,D_x)$ fulfills a global subelliptic estimate with loss of $\delta=2k_0/(2k_0+1)$ derivatives with respect to the elliptic case, that is, there exists $C>0$ such that for all $u \in D(q)$, 
$$\big\|\big(\langle(x,\xi)\rangle^{2/(2k_0+1)}\big)^w u\big\|_{L^2} \leq C\big(\|q^w(x,D_x) u\|_{L^2}+\|u\|_{L^2}\big),$$
with
$$D(q)=\{u \in L^2(\rr^n) : q^w(x,D_x)u \in L^2(\rr^n)\}.$$

By using the proof given for establishing Theorem~1.2.1 in \cite{karel},  one can actually directly obtain the more general estimate that there exists a constant $C>0$ such that for all $u \in D(q)$ and $\nu \in \rr$,  
\begin{equation}\label{est2}
\big\|\big(\langle(x,\xi)\rangle^{2/(2k_0+1)}\big)^w u\big\|_{L^2} \leq C\big(\|q^w(x,D_x) u-i\nu u\|_{L^2}+\|u\|_{L^2}\big).
\end{equation}
To prove this fact, it is actually sufficient to notice that one can replace the operator $q^{\textrm{Wick}}$ by the operator $q^{\textrm{Wick}}-i\nu$ in the estimate (2.10) in~\cite{karel}. 
Indeed, we recall\footnote{see the appendix about Wick calculus in~\cite{karel}} that real Hamiltonians get quantized in the Wick quantization by formally selfadjoint operators on $L^2$. Since the weight function $g$ given by Proposition~2.0.1 in~\cite{karel} is real-valued and $\eps$ a positive constant, we have the equality
$$\textrm{Re}\big((1-\eps g)^{\textrm{Wick}}(q^{\textrm{Wick}}-i\nu)\big)=\textrm{Re}\big((1-\eps g)^{\textrm{Wick}}q^{\textrm{Wick}}\big),$$
which justifies that one can actually replace the operator $q^{\textrm{Wick}}$ by the operator $q^{\textrm{Wick}}-i\nu$ in the estimate (2.10) in \cite{karel}. Then, by noticing that 
$$\textrm{Re}(q^{\textrm{Wick}}-i\nu)=\textrm{Re}(q^{\textrm{Wick}}),$$
one can also replace the operator $q^{\textrm{Wick}}$ by the operator $q^{\textrm{Wick}}-i\nu$ in the estimate preceding the estimate (2.19) in~\cite{karel}. Finally, after having done these two slight modifications, one can then use exactly the same proof as the one given for proving Theorem~1.2.1 in~\cite{karel} in order to establish the estimate (\ref{est2}).

We now need to recall few facts about the relationship between pseudodifferential calculus and functional calculus when using the Sobolev scale $\Lambda^r$, $r \in \rr$, defined by the operator
$$\Lambda^2=(1+|x|^2+|\xi|^2)^w=1+|x|^2+|D_{x}|^2.$$
In order to do so, it will be convenient to directly refer to the presentation given by F.~H\'erau and F.~Nier in \cite{HerNie02IHA} (Appendix~A, p.205). Our study corresponds to the specific case when the potential $V$ in~\cite{HerNie02IHA} is taken quadratic
$$V(x)=\frac{1}{2}|x|^2.$$ 
Define 
$$\mathbb{H}^r=\big\{u \in \mathscr{S}'(\rr^n) : (\langle (x,\xi)\rangle^r)^wu \in L^2(\rr^n)\big\}, \ r \in \rr$$
and
$$\|u\|_{\mathbb{H}^r}=\|(\langle (x,\xi)\rangle^r)^wu\|_{L^2}.$$
By referring to~\cite{HerNie02IHA} (Proposition~A.4), we notice that, for any $r \in \rr$, the operator $\Lambda^r$ is a pseudo-differential operator whose Weyl symbol belongs to the symbol class $S(\langle (x,\xi)\rangle^r,dx^2+d\xi^2)$ composed by $C^{\infty}(\rr_{x,\xi}^{2n},\cc)$ functions satisfying
$$\forall \alpha \in \nn^{2n},  \exists C_{\alpha}>0, \forall (x,\xi) \in \rr^{2n}, \ |\partial_{x,\xi}^{\alpha}a(x,\xi)| \leq C_{\alpha}\langle (x,\xi) \rangle^r.$$
Furthermore, for any $r \in \rr$, the domain of the operator $\Lambda^r$ is equal to the space $\mathbb{H}^r$ with equivalence of the two norms $\|u\|_{\mathbb{H}^r}$ and $\|\Lambda^r u \|_{L^2}$. It follows from (\ref{est2}) that there exists a constant $C>0$ such that for all $u \in D(q)$ and $\nu \in \rr$,  
\begin{equation}\label{est3}
\|\Lambda^{2/(2k_0+1)} u\|_{L^2}^2 \leq C\big(\|q^w(x,D_x) u-i\nu u\|_{L^2}^2+\|u\|_{L^2}^2\big).
\end{equation}
Consider now the metric
$$g=\frac{dx^2+d\xi^2}{\langle (x,\xi)\rangle^2}$$
and weight $M(x,\xi)=\langle (x,\xi)\rangle^2$.
We recall for instance from \cite{Le} (Lemma~2.2.18 and section~2.2) that this metric is admissible with gain $M$. Any quadratic form is obviously a first order symbol in the symbolic calculus associated to this metric
$$q \in S(M,g).$$ 
One can then directly deduce from the Fefferman-Phong inequality (See for instance Theorem~2.5.5 in~\cite{Le}) that there exists a positive $C>0$ such that for all $u \in D(\Lambda^2)$,
\begin{equation}\label{est4}
\|q^w(x,D_x)u\|_{L^2}^2 \leq C \|\Lambda^2 u\|_{L^2}^2,
\end{equation}
with 
$$D(\Lambda^2)=\{u \in L^2(\rr^n) : (1+|x|^2+|D_{x}|^2)u \in L^2(\rr^n)\}.$$
Notice from Theorem~\ref{th1} together with (\ref{vp0}) and (\ref{mi1}) that 
$$\sigma(q^w(x,D_x)) \cap i\rr=\sigma_{\textrm{disc}}\big(q^w(x,D_x)\big) \cap i\rr =\emptyset.$$
One can apply the abstract functional analysis led in \cite[Sec. 6.1]{HelNie05HES} in order to obtain a specific control of the resolvent of the operator $q^w(x,D_x)$ in particular regions of the resolvent set. More specifically, following the analysis in~\cite[p.67-69]{HelNie05HES}, we deduce from the hypoelliptic estimates (\ref{est3}) and (\ref{est4}) that there exists some positive constants $c$ and $C$ such that 
\begin{equation}\label{y1}
\{z \in \cc : \textrm{Re }z \geq -1/2, \ \textrm{Re }z +1 \leq c|z+1|^{\frac{1}{2k_0+1}}\} \cap \sigma(q^w(x,D_x))=\emptyset
\end{equation}
and
\begin{equation}\label{dl3}
\|(q^w(x,D_x)-z)^{-1}\| \leq C |z+1|^{-\frac{1}{2k_0+1}},
\end{equation}
for all $z \in \cc$ such that $\textrm{Re }z \geq -1/2$ and $\textrm{Re }z +1 \leq c|z+1|^{\frac{1}{2k_0+1}}$.

The property of exponential return to equilibrium may deduced from those resolvent estimates.
Indeed, Theorem~\ref{th2} is a consequence of the abstract analysis led  in~\cite[Theorem~6.1]{HelNie05HES} or~\cite{HeSjSt} (section~12, p.754-756). Following more specifically the presentation given in~\cite{HeSjSt} (section~12), we notice from (\ref{y1}) and (\ref{dl3}) that under the assumptions of Theorem~\ref{th2}, the quadratic operator $q^w(x,D_x)$ is a closed densely defined unbounded operator acting on $L^2(\rr^n)$ fulfilling the assumptions (12.1) and (12.2) in~\cite{HeSjSt} (p.754) with 
$$\delta=\frac{1}{2k_0+1}>0.$$
Define for any $t>0$,
$$E(t)=\frac{1}{2\pi i} \int_{\gamma}e^{-tz}\big(z-q^w(x,D_x)\big)^{-1}dz,$$
where $\gamma$ is a contour to the left of the spectrum that outside a compact set coincides with the curve
$$\textrm{Re }z=\frac{1}{C}|\textrm{Im }z|^{1/(2k_0+1)},$$
with $C>0$ and $\gamma$ oriented in the direction of decreasing $\textrm{Im }z$. According to (\ref{y1}) and (\ref{dl3}), this integral is convergent and defines a bounded operator depending smoothly on the parameter $t>0$. It is then proved in~\cite{HeSjSt} that this integral operator is actually the semigroup $e^{-tq^w(x,D_x)}$ generated by the accretive quadratic operator $q^w(x,D_x)$,
$$e^{-tq^w(x,D_x)}=\frac{1}{2\pi i} \int_{\gamma}e^{-tz}\big(z-q^w(x,D_x)\big)^{-1}dz.$$
Let $0 \leq \tau <\tau_0$ with $\tau_0$ the positive rate given by Theorem~\ref{th1}.
Keeping on following the analysis led in~\cite{HeSjSt} (p.755) in the particular case when $h=1$, we introduce two contours $\gamma$ and $\tilde{\gamma}$. 
\begin{figure}[ht]
\caption{}
\centerline{\includegraphics[scale=0.75]{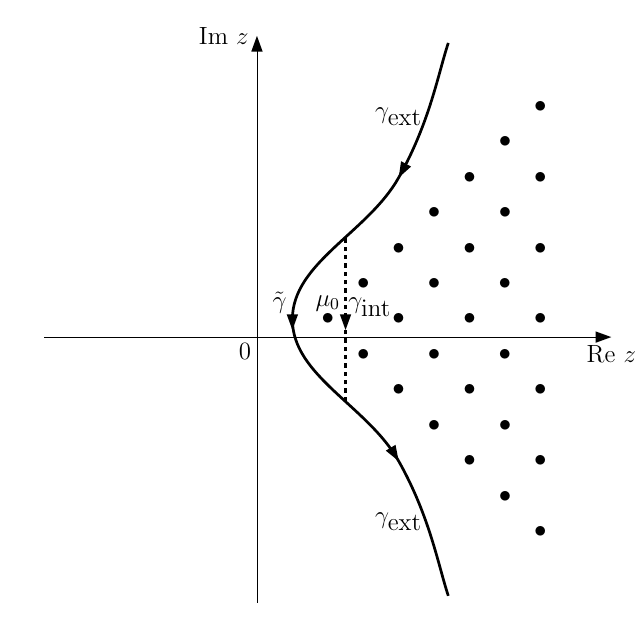}}
\end{figure}
Both contours are given by the curve
$$\textrm{Re }z=\frac{1}{C}|\textrm{Im }z|^{1/(2k_0+1)},$$
in the region where $\textrm{Re }z>b$, with $b=\textrm{Re }\mu_0+\tau$.
In the region, where $\textrm{Re }z \leq b$, the contour $\gamma$ is given by the equation $\textrm{Re }z=b$, while $\tilde{\gamma}$ joins the two points 
$$b+iC^{2k_0+1}b^{2k_0+1} \textrm{ and } b-iC^{2k_0+1}b^{2k_0+1},$$
further to the left so that $\tilde{\gamma}$ is entirely to the left of the spectrum of the quadratic operator $q^w(x,D_x)$, while $\gamma$ will only have the first eigenvalue in the bottom of the spectrum $\mu_0$ given by Theorem~\ref{th1} to its left, since $0 \leq \tau <\tau_0$. Let $\gamma_{\textrm{int}}$ denote the vertical part of the contour $\gamma$ in the region where $\textrm{Re }z=b$ and let $\gamma_{\textrm{ext}}$ denote the part of $\gamma$ in the region where $\textrm{Re }z \geq b$. 
On the exterior piece $\gamma_{\textrm{ext}}$, we have the resolvent estimate
\begin{equation}\label{y3}
\big\|\big(z-q^w(x,D_x)\big)^{-1}\big\| \leq \mathcal{O}\big(|\textrm{Im }z|^{-\frac{1}{2k_0+1}}\big),
\end{equation}
whereas on the interior piece $\gamma_{\textrm{int}}$, we have
\begin{equation}\label{y4}
\big\|\big(z-q^w(x,D_x)\big)^{-1}\big\| \leq \mathcal{O}(1),
\end{equation}
since the compact contour $\gamma_{\textrm{int}}$ does not intersect the spectrum of the operator $q^w(x,D_x)$.
According to Theorem~\ref{th1}, $\mu_0$ is an eigenvalue with algebraic multiplicity 1 for the operator $q^w(x,D_x)$.
Arguing as in (12.13) in the article~\cite{HeSjSt}, we may write that 
\begin{align}\label{y5}
e^{-tq^w(x,D_x)}=& \ \frac{1}{2\pi i} \int_{\tilde{\gamma}}e^{-tz}\big(z-q^w(x,D_x)\big)^{-1}dz\\ \label{y6}
=& \ e^{-\mu_0 t}\tilde{\Pi}_{\mu_0}+\frac{1}{2\pi i} \int_{\gamma}e^{-tz}\big(z-q^w(x,D_x)\big)^{-1}dz,
\end{align}
where $\tilde{\Pi}_{\mu_0}$ stands for the rank-one spectral projection associated with the eigenvalue $\mu_0$ for the operator $q^w(x,D_x)$. Using the decomposition $\gamma=\gamma_{\textrm{int}} \cup \gamma_{\textrm{ext}}$ together with (\ref{y3}) and (\ref{y4}), one can then estimate from the above the two parts of the integral
\begin{multline*}
\frac{1}{2\pi i} \int_{\gamma}e^{-tz}\big(z-q^w(x,D_x)\big)^{-1}dz=\frac{1}{2\pi i} \int_{\gamma_{\textrm{int}}}e^{-tz}\big(z-q^w(x,D_x)\big)^{-1}dz\\ +\frac{1}{2\pi i} \int_{\gamma_{\textrm{ext}}}e^{-tz}\big(z-q^w(x,D_x)\big)^{-1}dz,
\end{multline*}
as follows
$$\frac{1}{2\pi i} \int_{\gamma_{\textrm{int}}}e^{-tz}\big(z-q^w(x,D_x)\big)^{-1}dz=\mathcal{O}_{\mathcal{L}(L^2)}(e^{-bt}),$$
\begin{multline*}
\frac{1}{2\pi i} \int_{\gamma_{\textrm{ext}}}e^{-tz}\big(z-q^w(x,D_x)\big)^{-1}dz=\mathcal{O}_{\mathcal{L}(L^2)}(1) \int_{C^{2k_0+1}b^{2k_0+1}}^{+\infty}\hspace{-1cm}e^{-\frac{t}{C}y^{\frac{1}{2k_0+1}}}y^{-\frac{1}{2k_0+1}}dy\\
=\mathcal{O}_{\mathcal{L}(L^2)}(t^{-2k_0}) \int_{bt}^{+\infty}e^{-x}x^{2k_0-1}dx=\mathcal{O}_{\mathcal{L}(L^2)}(1)\Big(1+\frac{1}{t}+...+\frac{1}{t^{2k_0}}\Big)e^{-bt},
\end{multline*}
uniformly for $t \geq 1$.
Since the accretive operator $q^w(x,D_x)$ defines a contraction semigroup, we deduce from (\ref{y5}), (\ref{y6}) and the previous estimates that 
\begin{equation}\label{y7}
e^{-tq^w(x,D_x)}=e^{-\mu_0 t}\tilde{\Pi}_{\mu_0}+\mathcal{O}_{\mathcal{L}(L^2)}(e^{-bt}),
\end{equation} 
uniformly for $t \geq 0$.
Recalling that $b=\textrm{Re }\mu_0+\tau$ and considering now the operator 
$$Q=q^w(x,D_x)-\mu_0,$$ 
we directly deduce from (\ref{y7}) that there exists a positive constant $C>0$ such that 
$$\forall t \geq 0, \ \|e^{-tQ}-\Pi_0\|_{\mathcal{L}(L^2)} \leq Ce^{-\tau t},$$ 
where $\Pi_0$ is the rank-one spectral projection associated with the simple eigenvalue zero of the operator $Q$.

\subsection{Proof of Theorem~\ref{th3}} 
Let $q^w(x,D_x)$ be a quadratic operator satisfying the assumptions of Theorem~\ref{th1}. Assume furthermore that this quadratic operator is real and satisfies (\ref{y11}). As noticed in the discussion preceding the statement of Theorem~\ref{th3}, the lowest eigenvalue $\mu_0$ is necessarily real and the quadratic form $a$ defining the ground state 
$$u_0(x)=e^{-a(x)} \in \mathscr{S}(\rr^n), \quad q^w(x,D_x)u_0=\mu_0 u_0,$$
is positive definite. It remains to check that the rank-one spectral projection $\Pi_0$ associated with the simple eigenvalue zero of the operator 
$$Q=q^w(x,D_x)-\mu_0,$$ 
provided by Theorem~\ref{th2} is orthogonal 
$$\Pi_0u=\|u_0\|_{L^2(\rr^n)}^{-2}(u,u_0)_{L^2(\rr^n)}u_0, \ u \in L^2(\rr^n),$$
in order to deduce Theorem~\ref{th3} from Theorem~\ref{th2}. This property is a direct consequence of the assumption
\begin{equation}\label{yy11}
\textrm{Ker}\big(q^w(x,D_x)-\mu_0\big)=\textrm{Ker}\big(q^w(x,D_x)^*-\mu_0\big)=\cc u_0 \subset \mathscr{S}(\rr^n).
\end{equation}
Indeed, notice from (6.52) in~\cite[III.6]{kato} that the adjoint of the spectral projection $\Pi_0$ associated with the eigenvalue zero for the operator $Q$ 
is also a spectral projection for the operator $Q^*$,
\begin{equation}\label{y14}
\Pi_0=\frac{1}{2i\pi}\int_{\Gamma}(z-Q)^{-1}dz, \ \Pi_0^*=\frac{1}{2i\pi}\int_{\overline{\Gamma}}(z-Q^*)^{-1}dz,
\end{equation}
with $\overline{\Gamma}$ being the mirror image of $\Gamma$ with respect to the real axis and $\Gamma$ a circular contour centered in the eigenvalue $0$ with a sufficiently small radius. Notice that in the integrals (\ref{y14}) both contours $\Gamma$ and $\overline{\Gamma}$ are taken in the positive direction and that $\Pi_0^*$ is actually the spectral projection  associated with the eigenvalue zero for the operator $Q^*$.
Since $\Pi_0$ and $\Pi_0^*$ are both projections, one may decompose the $L^2(\rr^n)$ space in two (possibly different) direct sums
$$L^2(\rr^n)=\textrm{Ker}(\Pi_0)\oplus \textrm{Ran}(\Pi_0) \textrm{ and } L^2(\rr^n)=\textrm{Ker}(\Pi_0^*)\oplus \textrm{Ran}(\Pi_0^*).$$  
It follows from (\ref{yy11}) that 
$$ \textrm{Ran}(\Pi_0)=\textrm{Ran}(\Pi_0^*)=\cc u_0.$$
Indeed, Theorem~\ref{th1} applies for both quadratic operators $q^w(x,D_x)$ and $q^w(x,D_x)^*$ and the eigenvalue zero has therefore algebraic multiplicity 1 for both operators $Q$ and $Q^*$. 
Furthermore, since 
$$\textrm{Ker}(\Pi_0)=\textrm{Ran}(\Pi_0^*)^{\perp}=(\cc u_0)^{\perp} \textrm{ and } \textrm{Ker}(\Pi_0^*)=\textrm{Ran}(\Pi_0)^{\perp}=(\cc u_0)^{\perp},$$ 
we conclude that the spectral projection $\Pi_0$ is orthogonal
$$\Pi_0u=\|u_0\|_{L^2(\rr^n)}^{-2}(u,u_0)_{L^2(\rr^n)}u_0, \ u \in L^2(\rr^n).$$  
This ends the proof of Theorem~\ref{th3}.

%
%
\section {Some applications}\label{applications}

\noindent
In this section, several applications of the singular space theory are discussed. In particular, in addition to the standard Fokker-Planck operator, we consider the spectral problem for the Fokker-Planck (forward Kolmogorov) operator of two Markovian stochastic systems that appear in non-equilibrium statistical mechanics. The starting point is that of a classical open system, i.e. of a `small' Hamiltonian system (the distinguished particle) coupled to one or more heat baths at different temperatures which are modelled as linear wave equations with initial conditions distributed according to appropriate Gibbs  measures \cite{Rey-Bellet2006}. After eliminating the heat bath variables one obtains a non-Markovian evolution equation describing the dynamics of the distinguished particle. A Markovian approximation of this non-Markovian dynamics leads to a system of stochastic differential equations that is quite similar to the Langevin equation~\eqref{e:langevin}, see equations~\eqref{system1} and~\eqref{osc1} below. In fact, the Langevin dynamics~\eqref{e:langevin} and the corresponding Fokker-Planck equation~\eqref{e:fp} can be obtained from the dynamics~\eqref{system1} in the limit of rapid decorrelation of the noise. For further details, see~\cite{OttobrePavliotis09} and the references therein.   

In Section~\ref{subsec:fp} we revisit the spectral problem of the Fokker-Planck operator with a quadratic potential using the techniques developed in this paper. In Section~\ref{subsec:GLE} we study the Fokker-Planck operator for the Markovian approximation of the generalized Langevin equation (GLE) with a quadratic potential. In Section~\ref{subsec:chain} we investigate the spectrum of the Fokker-Planck operator for a chain of oscillators coupled to two heat baths with different temperature, under the assumption of quadratic confining as well as quadratic interaction potentials. Even though the Fokker-Planck operators, or equivalently the dynamics~\eqref{system1} and~\eqref{osc1}, that we study in Sections~\ref{subsec:GLE} and~\ref{subsec:chain} are quite similar, we choose to analyze them separately. The main reason for this is that, unlike the GLE case, in the chain of oscillators problem it is not possible to write down a simple explicit formula for the ground state; this fact complicates the analysis. We believe that the examples presented in this section illustrate the fact that quadratic hypoelliptic operators with nonnegative real parts and zero singular spaces behave essentially like the Kramers-Fokker-Planck operator.

\subsection{The Kramers-Fokker-Planck operator with quadratic potential}
\label{subsec:fp}
As pointed out in~\cite{hps},
a noticeable example of quadratic operator with zero singular space is the Kramers-Fokker-Planck operator considered on the unweighted $L^2(\rr_{x,v}^2)$ space
$$K=-\Delta_v+\frac{v^2}{4}-\frac{1}{2}+v.\partial_x-\nabla_xV(x).\partial_v, \ (x,v) \in \rr^{2},$$
with a quadratic potential
$$V(x)=\frac{1}{2}ax^2, \ a \in \rr^*.$$
Indeed, this operator writes as 
$$K=q^w(x,v,D_x,D_v)-\frac{1}{2},$$
with a symbol
$$q(x,v,\xi,\eta)=\eta^2+\frac{1}{4}v^2+i(v \xi-a x \eta),$$
which is a non-elliptic complex-valued quadratic form whose real part is non-negative.
Simple algebraic computations show that the associated Hamilton map 
$$q(x,v,\xi,\eta)=\sigma\big((x,v,\xi,\eta),F(x,v,\xi,\eta) \big),$$
is 
$$F= \left( \begin{array}{cccc}
  0 & \frac{1}{2}i & 0 & 0 \\
  -\frac{1}{2}ai& 0 & 0 & 1 \\
0 & 0 & 0 &  \frac{1}{2}ai \\
0 & -\frac{1}{4} & -\frac{1}{2}i & 0 
  \end{array}
\right),$$
and that the singular space is equal to zero $S=\{0\}$. More precisely, the integer $k_0$ appearing in (\ref{er133}) is equal to~1:
$$\textrm{Ker}(\textrm{Re }F) \cap \textrm{Ker}(\textrm{Re }F \ \textrm{Im }F) \cap \rr^{4}=\{0\}.$$
Now a computation shows that $\lambda \in \cc$ is an eigenvalue of $2F$
precisely when
$$
\left(\lambda-\frac{a}{\lambda}\right)^2 + 1 =0,
$$
and we easily see that when $a<0$, the eigenvalues $\lambda$ of $2F$ with $\textrm{Im }\lambda>0$ are given by
$$
\lambda_1 = \frac{-i+i\sqrt{1-4a}}{2},\quad \lambda_2 = \frac{i+i\sqrt{1-4a}}{2}.
$$
When $a>0$, we get the eigenvalues
$$
\lambda_1  = \frac{i+i\sqrt{1-4a}}{2},\quad \lambda_2 = \frac{i-i\sqrt{1-4a}}{2}.
$$
According to (\ref{sm6}), the spectrum of the operator $q^w(x,v,D_x,D_v)$ is given by
$$
\left\{
\left(\frac{1}{2}+k_1\right) \frac{\lambda_1}{i}+\left(\frac{1}{2}+k_2\right)\frac{\lambda_2}{i},\quad k_j \in \nn \right\}.
$$
In particular, when $a>0$, we observe that the lowest eigenvalue of the spectrum of the operator $q^w(x,v,D_x,D_v)$ is 
$$\mu_0=\frac{1}{2},$$ 
whereas when $a<0$, this lowest eigenvalue is equal to 
$$\mu_0=\frac{\sqrt{1-4a}}{2}.$$
The spectral gap $\tau_0>0$ appearing in Theorem~\ref{th1} is respectively equal to 
$$\frac{\sqrt{1-4a}-1}{2}, \quad \frac{1-\sqrt{1-4a}}{2}, \quad \frac{1}{2},$$ 
when 
$$a<0, \quad 0<a \leq 1/4, \quad a>1/4.$$ 
Notice that the lowest eigenvalue is always real. This is consistent with the fact that the quadratic operator $q^w(x,v,D_x,D_v)$ is real. Further calculations allow to determine explicitly the ground state
$$q^w(x,v,D_x,D_v)u_0=\mu_0 u_0.$$
When $a>0$, we obtain the usual Maxwellian 
$$u_0(x,v)=e^{-\frac{1}{4}ax^2-\frac{v^2}{4}}=e^{-\frac{1}{2}(\frac{v^2}{2}+V(x))} \in \mathscr{S}(\rr^2),$$
whereas when $a<0$, the ground state is given by
$$u_0(x,v)=e^{\frac{a}{4}\sqrt{1-4a}x^2-axv-\frac{\sqrt{1-4a}}{4}v^2} \in \mathscr{S}(\rr^2).$$
Notice further that the assumption 
$$\textrm{Ker}\big(q^w(x,D_x)-\mu_0\big)=\textrm{Ker}\big(q^w(x,D_x)^*-\mu_0\big)=\cc u_0 \subset \mathscr{S}(\rr^2),$$
holds true only if $a>0$. Theorems~\ref{th1} and \ref{th2} therefore apply whenever $a \neq 0$, whereas Theorem~\ref{th3} only applies when $a>0$.
%
%
\subsection{The generalized Langevin equation}
\label{subsec:GLE}

The spectrum of the Fokker-Planck operator associated to a linear stochastic differential equation in $\rr^{d}$ has been calculated in~\cite{Metafune_al2002}, see also~\cite{Lorenzi}. In this section we consider a particular system of linear stochastic differential equations that is obtained as a finite-dimensional Markovian approximation of the non-Markovian generalized Langevin equation (GLE) in $\rr^d$,
\begin{equation}\label{GLE}
\ddot{x}=-\nabla_xV(x)-\int_0^t \gamma(t-s)\dot{x}(s)ds+F(t),
\end{equation}
where $V(x)$ is a smooth confining potential and $F(t)$ a mean zero stationary Gaussian process with autocorrelation function $\gamma(t)$, in accordance to the fluctuation-dissipation theorem
$$\langle F(t) \otimes F(s) \rangle=\beta^{-1}\gamma(t-s)I.$$
Here $\beta>0$ stands for the inverse temperature and $I$ for the identity matrix. The GLE equation together with the fluctuation-dissipation theorem appear in various settings such as surface diffusion \cite{rreff1} and polymer dynamics \cite{rreff2}.  

For simplicity, we will consider the case $d=1$. As in~\cite{OttobrePavliotis09}, we study the case when the memory kernel is given by a sum of exponentials 
$$\gamma_m(t)=\sum_{j=1}^m\lambda_j^2 e^{-\alpha_j\vert t\vert} ,$$ with $\alpha_j>0$, $\lambda_j \neq 0$, $j=1,...,m$. Equation (\ref{GLE}) is then equivalent to the following dynamics 
\begin{equation}\label{system1}
\left\lbrace
  \begin{array}{l}
dx=ydt,\\
dy=-\nabla_xV(x)dt+\sum_{j=1}^m\lambda_jz_jdt,\\
dz_j=-\left(\lambda_j y+\alpha_jz_j\right)dt+\sqrt{2\alpha_j \beta^{-1}}  \,dW_j,
\quad j=1,...,m,
\end{array}
\right.
\end{equation}
with $(x,y,z)\in \rr^{m+2}$, $z=(z_1,...,z_m) \in \rr^{m}$ and $(W_j)_{j=1...m}$ being independent standard Brownian motions. 

We consider the specific case when $V$ is a non-degenerate quadratic potential 
$$V(x)=\frac{1}{2} \omega^2x^2,$$
with $\omega \neq 0$.
The density of the invariant measure associated to the dynamics (\ref{system1}) is then given by
$$
\rho(x,y,z)=\frac{1}{Z_{\beta}}e^{-\beta(V(x)+\frac{y^2}{2}+
\frac{|z|^2}{2})},
$$ 
with $Z_{\beta}$ being the normalization constant and $|\cdot|$ the Euclidean norm on $\rr^m$. The dynamics~\eqref{system1} is Markovian whose generator
\begin{equation}\label{generator}
\mathcal{L}=y \partial_x-V'(x) \partial_y+\Big(\sum_{j=1}^m \lambda_jz_j\Big) 
\partial_y -y \sum_{j=1}^m \lambda_j\partial_{z_j}-
\sum_{j=1}^m \alpha_jz_j\partial_{z_j}+\sum_{j=1}^m\alpha_j \beta^{-1} \partial_{z_j}^2,
\end{equation}
is a hypoelliptic operator acting on $L^2_{\rho}=L^2(\rho dxdydz)$, with $L^2$-adjoint 
$$
\mathcal{L}^*=-y \partial_x+V'(x)\partial_y-\Big(\sum_{j=1}^m \lambda_jz_j\Big)
\partial_y +y\sum_{j=1}^m \lambda_j\partial_{z_j}+
\sum_{j=1}^m \alpha_j\partial_{z_j}(z_j\cdot)+\sum_{j=1}^m\alpha_j \beta^{-1}\partial_{z_j}^2,
$$
acting on $L^2_{\rho^{-1}}$. By using the transformation 
$$\mathcal{H}=\rho^{-1/2}\mathcal{L}^*(\rho^{1/2}\cdot),$$
in order to work with an operator acting on the unweighted $L^2(\rr_{x,y,z}^{m+2})$ space
$$\mathcal{H}=-y\partial_x+\omega^2x\partial_y-\Big(\sum_{j=1}^m \lambda_jz_j\Big)
\partial_y +y\sum_{j=1}^m \lambda_j\partial_{z_j}-
\sum_{j=1}^m \alpha_j \Big(-\beta^{-1}\partial_{z_j}^2+\frac{\beta}{4}z_j^2\Big) 
+\frac{1}{2}\sum_{j=1}^m \alpha_j$$
and setting
\begin{equation}\label{H}
\tilde{\mathcal{H}}=-\mathcal{H}+\frac{1}{2}\sum_{j=1}^m \alpha_j,
\end{equation}
we notice that 
\begin{multline*}
\tilde{\mathcal{H}}=q^w(x,y,z,D_x,D_y,D_z)=y\partial_x-\omega^2x\partial_y+\Big(\sum_{j=1}^m \lambda_jz_j\Big)
\partial_y \\ -y\sum_{j=1}^m \lambda_j\partial_{z_j}+
\sum_{j=1}^m \alpha_j\Big(-\beta^{-1}\partial_{z_j}^2+\frac{\beta}{4} z_j^2\Big),
\end{multline*}
is a quadratic operator, whose symbol is given by  
$$q(x,y,z,\xi,\eta,\zeta)=i(y\xi-\omega^2x\eta)+ i\sum_{j=1}^m\lambda_j(z_j\eta
-y\zeta_j)+\sum_{j=1}^m\alpha_j \left( \beta^{-1} \zeta_j^2+\beta\frac{z_j^2}{4} \right).$$ 
This quadratic symbol has a non-negative real part $\textrm{Re }q\geq 0$ and a direct computation shows that its Hamilton map
$$q(x,y,z,\xi,\eta,\zeta)=\sigma\big((x,y,z,\xi,\eta,\zeta),F(x,y,z,\xi,\eta,\zeta)\big),$$
is given by $(\tilde{x},\tilde{y},\tilde{z},\tilde{\xi},\tilde{\eta},\tilde{\zeta})=F(x,y,z,\xi,\eta,\zeta)$, with 
$$\tilde{x}=\frac{1}{2}iy, \ \tilde{y}=-\frac{1}{2}iw^2x+\frac{1}{2}i\sum_{j=1}^m\lambda_j z_j, \ \tilde{z}_j=-\frac{1}{2}i \lambda_j y+\alpha_j \beta^{-1} \zeta_j, \ \tilde{\xi}=\frac{1}{2}i w^2 \eta,$$
$$\tilde{\eta}=-\frac{1}{2}i \xi+\frac{1}{2}i \sum_{j=1}^m\lambda_j \zeta_j, \ \tilde{\zeta}_j=-\frac{1}{2}i\lambda_j \eta-\frac{1}{4}\beta \alpha_j z_j.$$
It directly follows that
$$\textrm{Ker}(\textrm{Re } F)\cap \R^{2(m+2)}=\{(x,y,z,\xi,\eta,\zeta)\in
\R^{2(m+2)}: z=\zeta=0\},$$
\begin{multline*}
\textrm{Ker}(\textrm{Re } F)\cap \textrm{Ker}(\textrm{Re }F
\textrm{Im }F)\cap \R^{2(m+2)}\\ =\{(x,y,z,\xi,\eta,\zeta)\in
\R^{2(m+2)}: y=z=\eta=\zeta=0\},
\end{multline*}
\begin{multline*}
\textrm{Ker}(\textrm{Re } F)\cap \textrm{Ker}(\textrm{Re }F
\textrm{Im }F)\cap\textrm{Ker}\big(\textrm{Re }F
(\textrm{Im }F)^2\big)\cap \R^{2(m+2)}\\ =\{(x,y,z,\xi,\eta,\zeta)\in
\R^{2(m+2)}: x=y=z=\xi=\eta=\zeta=0\}.
\end{multline*}
This proves that the singular space associated to the symbol $q$ reduces to $\{0\}$ after intersecting exactly $k_0+1$ kernels with here $0 \leq k_0=2 \leq 2(m+2)-1$.
Furthermore, notice that the quadratic operator $\mathcal{H}=q^w(x,y,z,D_x,D_y,D_z)$ is real. Setting 
$$u_0(x,y,z)=Z_{\beta}^{1/2}\rho(x,y,z)^{1/2}=e^{-\frac{\beta}{2}(V(x)+\frac{y^2}{2}+
\frac{|z|^2}{2})}=e^{-\frac{\beta}{4}(w^2x^2+y^2+|z|^2)} \in \mathscr{S}(\rr^{m+2})$$
and 
$$\mu_0=\frac{1}{2}\sum_{j=1}^m \alpha_j >0,$$
a direct computation shows 
$$\tilde{\mathcal{H}}u_0=\tilde{\mathcal{H}}^*u_0=\mu_0 u_0.$$
On the other hand, notice that
\begin{multline*}
\textrm{Re}(\tilde{\mathcal{H}}u,u)_{L^2}=\textrm{Re}(\tilde{\mathcal{H}}^*u,u)_{L^2}=\sum_{j=1}^m \alpha_j\Big(\|\beta^{-1/2}D_{z_j}u\|_{L^2}^2+ \Big\|\frac{\beta^{1/2}}{2}z_ju\Big\|_{L^2}^2\Big) \\ \geq 
2\sum_{j=1}^m \alpha_j\|\beta^{-1/2}D_{z_j}u\|_{L^2}\Big\|\frac{\beta^{1/2}}{2}z_ju\Big\|_{L^2}  \geq \sum_{j=1}^m \alpha_j\|D_{z_j}u\|_{L^2}\|z_ju\|_{L^2} \geq \mu_0\|u\|_{L^2}^2,
\end{multline*}
since
$$\sum_{j=1}^m \alpha_j\|D_{z_j}u\|_{L^2}\|z_ju\|_{L^2} \geq \sum_{j=1}^m \alpha_j\textrm{Re}(D_{z_j}u,iz_ju)_{L^2}=\frac{1}{2} \sum_{j=1}^m \alpha_j([D_{z_j},iz_j]u,u)_{L^2}$$
and
$$\frac{1}{2}\sum_{j=1}^m \alpha_j([D_{z_j},iz_j]u,u)_{L^2}=\frac{1}{2}\sum_{j=1}^m \alpha_j\|u\|_{L^2}^2=\mu_0\|u\|_{L^2}^2.$$
Here the notation $[D_{z_j},iz_j]$ stands for the commutator of the two operators $D_{z_j}=i^{-1}\partial_{z_j}$ and $iz_j$.
This implies that $\mu_0$ is necessarily the lowest eigenvalue in the spectrum of the quadratic operator $\tilde{\mathcal{H}}$ and that $u_0$ is the associated ground state appearing in Theorem~\ref{th1}. We notice further that all the assumptions of Theorems~\ref{th1}, \ref{th2} and \ref{th3} hold true. Thus, these three theorems apply. This proves the property of exponential return to equilibrium for this finite-dimensional Markovian approximation of the non-Markovian generalized Langevin equation with  quadratic potential. Here, Theorem~\ref{th1} provides an exact formula for the rate of convergence that may be computed explicitly
$$\tau_0=2 \min_{\substack{\lambda \in \sigma(F) \\   \textrm{Im }\lambda>0}} \textrm{Im }\lambda.$$
For simplicity, consider the specific case when $m=1$ and denote $\lambda$ and $\alpha$ for the parameters $\lambda_1$ and $\alpha_1$. An explicit computation of the characteristic polynomial of $F$ shows that
$$P(X)=\det(F-X\textrm{I})=X^6+\frac{1}{4}(\alpha^2-2\lambda^2-2w^2)X^4+\frac{1}{16}\big((\lambda^2+w^2)^2-2w^2 \alpha^2\big)X^2+\frac{w^4\alpha^2}{64}.$$
This is a polynomial equation of the third degree in the variable $Y=X^2$ which may be solved explicitly. Refraining from using Cardan formulas for giving an explicit, but rather complicated, formula for the rate of convergence $\tau_0$, we may instead try to compute numerically this rate in order to study its dependence on the parameters $\alpha$ and $\lambda$. A limited attempt in this direction is provided by the following calculations where the rate $\tau_0$ is numerically computed for different values of the parameters  $\lambda$ and $\alpha$ in the case when $w=1$. Notice in particular that the spectral gap seems to be maximized along a curve of the type $\alpha=\gamma \lambda^2$, with $\gamma>0$ a positive constant. On the second figure, the spectral gap is computed as a function of the parameter $\lambda$ while keeping the friction coefficient $\gamma = \frac{\lambda^2}{\alpha}$ fixed. We observe that the spectral gap becomes almost constant for a large range of value
 s of $\lambda$. This is in accordance with the numerical observations reported in~\cite{CeriottiBussiParrinello2009}. It would be interesting to maximize the spectral gap, i.e. to optimize the rate of convergence to equilibrium, by choosing appropriately the parameters $\alpha_j$ and $\lambda_j$ in~\eqref{system1}. This question, together with possible applications to Markov Chain Monte Carlo (MCMC) techniques will be studied in future work.

\begin{figure}[ht]
\label{pic1}
\caption{Spectral gap $\tau_0$ as a function of $\alpha$ and $\lambda$.}
\begin{center}
\includegraphics[scale=0.45]{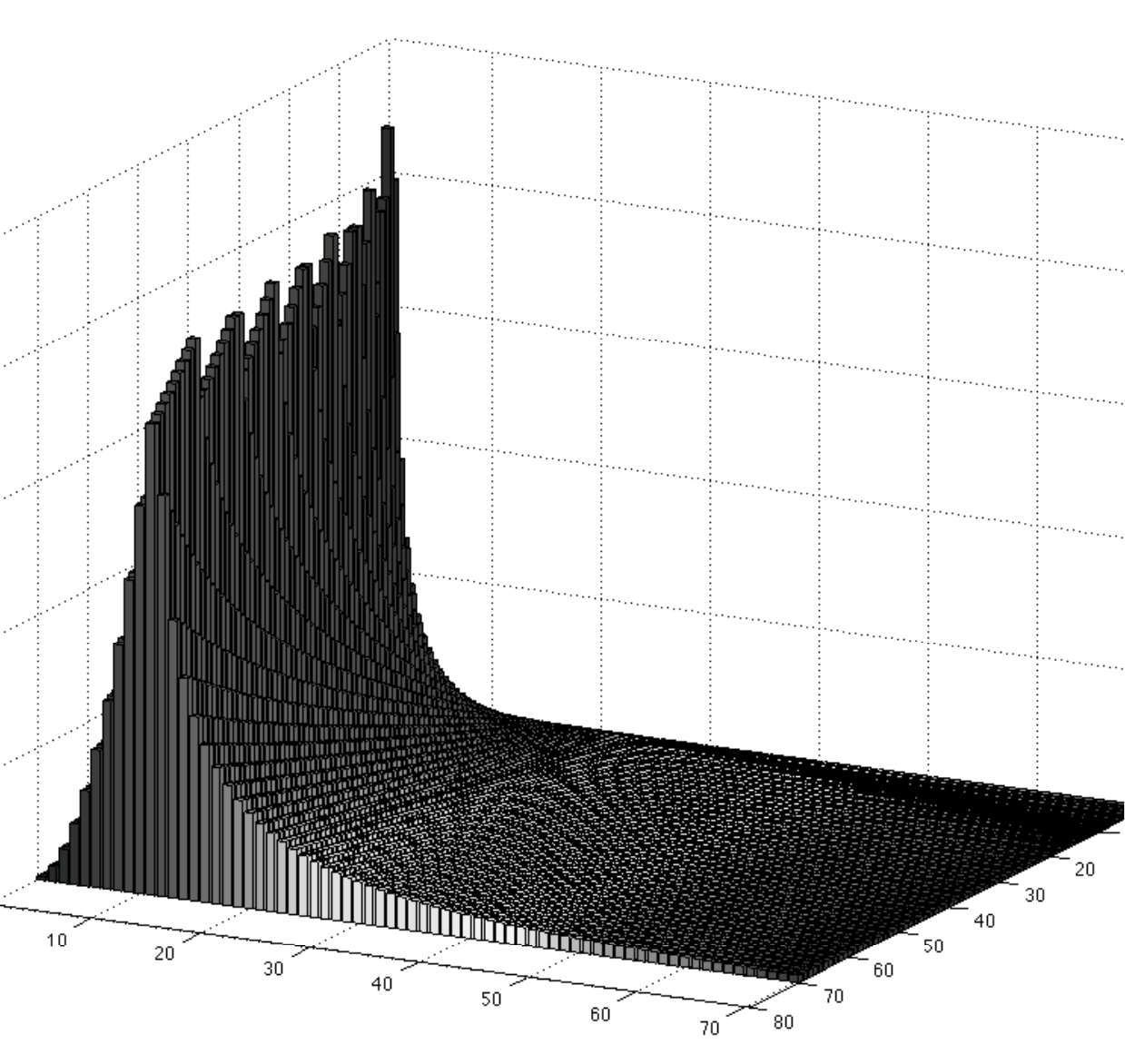}
\end{center}
\end{figure}
\begin{figure}[ht]
\caption{Spectral gap $\tau_0$ as a function of $\lambda$ with $\gamma = \frac{\lambda^2}{\alpha}$ fixed.}
\label{fig:spec_gap_gamma_fixed}
\begin{center}
\includegraphics[scale=0.45]{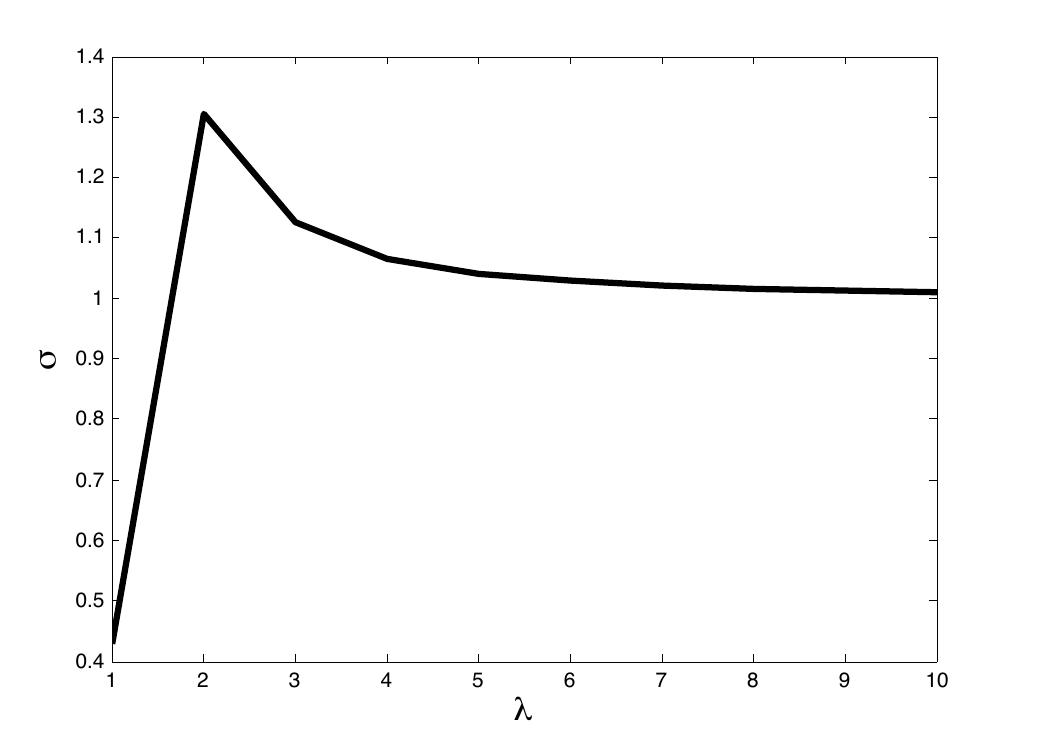}
\end{center}
\end{figure}
%
%
\subsection{Chain of oscillators}
\label{subsec:chain}
This application comes from the series of works \cite{EH00,EH03,EPR99} and was studied in~\cite{HeHiSj2} (Section~6). We shall begin by recalling the setting of the analysis~\cite{HeHiSj2} and explain how the results obtained there relate with the present work.

This example is a model describing a chain of two oscillators coupled with two heat baths at each side. The particles are described by their respective position and velocity $(x_j,y_j) \in \rr^{2d}$. For each oscillator $j \in \{1,2\}$, the particles are submitted to an external force derived from a real-valued potential $V_j(x_j)$ and a coupling between the two oscillators derived from a real-valued potential $V_c(x_2-x_1)$. We write $V$ the full potential
$$V(x)=V_1(x_1)+V_2(x_2)+V_c(x_2-x_1), \ x=(x_1,x_2) \in \rr^{2d},$$
$y=(y_1,y_2) \in \rr^{2d}$ the velocities and $z=(z_1,z_2) \in \rr^{2d}$ the variables describing the state of the particles in each of the heat baths. In each bath, the particles are submitted to a coupling with the nearest oscillator, a friction force given by the friction coefficient $\gamma$ and a thermal diffusion at temperature $T_j$, $j \in \{1,2\}$. We denote by $w_j$, $j \in \{1,2\}$, two standard $d$-dimensional Brownian motions and write $w=(w_1,w_2)$. The system of equations describing this model, and which is obtained from non-Markovian dynamics of the form~\eqref{GLE} through a similar Markovian approximation, reads as (see~\cite{EPR99}),
\begin{equation}\label{osc1}
\left\lbrace
  \begin{array}{l}
    dx_1 = y_1 dt\\
    dx_2=y_2dt\\
    dy_1 = -\partial_{x_1}V(x)dt+z_1dt\\
    dy_2 = -\partial_{x_2}V(x)dt+z_2dt\\
    dz_1=- \gamma z_1 dt+\gamma x_1 dt-\sqrt{2 \gamma T_1}dw_1\\
    dz_2=-\gamma z_2dt+\gamma x_2 dt-\sqrt{2\gamma T_2}dw_2.
  \end{array}
\right.
\end{equation}
Following~\cite{HeHiSj2} and setting $T_1=\alpha_1 h/2$, $T_2=\alpha_2 h/2$, the corresponding equation for the density of particles becomes 
\begin{multline}\label{osc2}
    h\partial_t f+\frac{\gamma}{2}\alpha_1 (-h\partial_{z_1})\Big(h\partial_{z_1}+\frac{2}{\alpha_1}(z_1-x_1)\Big)f\\ +\frac{\gamma}{2}\alpha_2(-h\partial_{z_2})\Big(h\partial_{z_2}+\frac{2}{\alpha_2}(z_2-x_2)\Big)f+\Big(yh\partial_x-(\nabla_xV(x)-z)h\partial_y\Big)f =0.
\end{multline}
Define
$$\Phi(x,y,z)=V(x)+\frac{y^2}{2}+\frac{z^2}{2}-zx$$
and
$$\mathcal{M}_{\alpha}=\frac{1}{C}e^{-\frac{2\Phi}{\alpha h}},$$
with $\alpha>0$ and $C>0$.
One may check that the function $\mathcal{M}_{\alpha}$, with $\alpha=\alpha_1=\alpha_2$, is a Maxwellian of the process when the temperatures are the same. Considering now the general case when temperatures may be different, the function $\mathcal{M}_{\alpha}$ is used to define the weighted space $L^2(e^{-\frac{2\Phi}{\alpha h}}dxdydz)$. In order to work in the flat $L^2$ space, we change the unknown by setting $f=\mathcal{M}_{\alpha}^{1/2}u$. The new equation for the unknown $u$ reads as
\begin{multline}\label{osc3}
h\partial_t u+\frac{\gamma}{2}\alpha_1 \Big(-h\partial_{z_1}+\frac{1}{\alpha}(z_1-x_1)\Big)\Big(h\partial_{z_1}+\big(\frac{2}{\alpha_1}-\frac{1}{\alpha}\Big)(z_1-x_1)\Big)u+\\ \frac{\gamma}{2}\alpha_2\Big(-h\partial_{z_2}+\frac{1}{\alpha}(z_2-x_2)\Big)\Big(h\partial_{z_2}+\Big(\frac{2}{\alpha_2}-\frac{1}{\alpha}\Big)(z_2-x_2)\Big)u+\Big(yh\partial_x-(\nabla_xV(x)-z)h\partial_y\Big)u =0.
\end{multline}
We consider the case when external potentials are quadratic. For simplicity, we may assume that $h=1$, $\gamma=2$, $d=1$ and take 
\begin{equation}\label{osc10}
V_1(x_1)=\frac{1}{2}ax_1^2, \quad V_2(x_2)=\frac{1}{2}b x_2^2, \quad V_c(x_1-x_2)=\frac{1}{2}c(x_1-x_2)^2,
\end{equation}
with $a,b,c \in \rr$. Equation (\ref{osc3}) writes as 
$$\partial_tu+q^{w}(X,D_X)u-2u=0, \ X=(x,y,z) \in \rr^{6},$$
where $q^{w}(X,D_X)$ is the quadratic operator with symbol
\begin{multline*}
q=\alpha_1\zeta_1^2+\alpha_2\zeta_2^2+\beta_1(z_1-x_1)^2+\beta_2(z_2-x_2)^2+i\big[2\delta_1\zeta_1(z_1-x_1)+2\delta_2\zeta_2(z_2-x_2)\\ +y_1\xi_1+y_2\xi_2-\eta_1\big((a+c)x_1-cx_2-z_1\big)-\eta_2\big(-cx_1+(b+c)x_2-z_2\big)\big],
\end{multline*}
with 
$$\beta_1=\frac{\alpha_1}{\alpha}\Big(\frac{2}{\alpha_1}-\frac{1}{\alpha}\Big), \ \beta_2=\frac{\alpha_2}{\alpha}\Big(\frac{2}{\alpha_2}-\frac{1}{\alpha}\Big), \ \delta_1=\frac{\alpha_1}{\alpha}-1, \ \delta_2=\frac{\alpha_2}{\alpha}-1,$$
the notations $\xi, \eta,\zeta$ standing respectively for the dual variables of $x,y,z$. The condition 
$$\alpha \geq \frac{1}{2}\max(\alpha_1,\alpha_2),$$
appearing in~\cite{HeHiSj2} (Section~6) exactly ensures that this quadratic symbol has a non-negative real part $\textrm{Re }q \geq 0$. Notice that only the case with identical temperatures $\alpha=\alpha_1=\alpha_2$ is discussed in~\cite{HeHiSj2}. More precisely, the authors mention that no Maxwellian is known in the case of different temperatures and that it prevents them from finding any supersymmetric structure. 
Here, we consider the general case with possibly different temperatures $\alpha_1 \neq \alpha_2$ for the quadratic potentials defined in (\ref{osc10}) and assume that  
$$\alpha > \frac{1}{2}\max(\alpha_1,\alpha_2), \ \alpha_1>0,\ \alpha_2>0.$$
Direct algebraic computations using the definition of Hamilton maps and the explicit expressions 
$$\textrm{Re }q=\alpha_1\zeta_1^2+\alpha_2\zeta_2^2+\beta_1(z_1-x_1)^2+\beta_2(z_2-x_2)^2,$$
\begin{multline*}
\textrm{Im }q= 2\delta_1\zeta_1(z_1-x_1)+2\delta_2\zeta_2(z_2-x_2)+y_1\xi_1+y_2\xi_2\\ -\eta_1\big((a+c)x_1-cx_2-z_1\big)-\eta_2\big(-cx_1+(b+c)x_2-z_2\big),
\end{multline*}
show that  $(\tilde{x},\tilde{y},\tilde{z},\tilde{\xi},\tilde{\eta},\tilde{\zeta})=(\textrm{Re }F)(x,y,z,\xi,\eta,\zeta)$ with
$$\tilde{x}_1=0, \ \tilde{x}_2=0, \ \tilde{y}_1=0, \ \tilde{y}_2=0, \tilde{z}_1=\alpha_1 \zeta_1,\ \tilde{z}_2=\alpha_2 \zeta_2, \ \tilde{\xi}_1=\beta_1(z_1-x_1),$$
$$\tilde{\xi}_2=\beta_2(z_2-x_2), \ \tilde{\eta}_1=0, \ \tilde{\eta}_2=0, \ \tilde{\zeta}_1=-\beta_1(z_1-x_1), \tilde{\zeta}_2=-\beta_2 (z_2-x_2),$$
and $(\tilde{x},\tilde{y},\tilde{z},\tilde{\xi},\tilde{\eta},\tilde{\zeta})=(\textrm{Im }F)(x,y,z,\xi,\eta,\zeta)$ with
$$\tilde{x}_1=\frac{1}{2}y_1, \ \tilde{x}_2=\frac{1}{2}y_2, \ \tilde{y}_1=-\frac{1}{2}\big((a+c)x_1-cx_2-z_1\big), \ \tilde{y}_2=-\frac{1}{2}\big(-cx_1+(b+c)x_2-z_2\big),$$
$$\tilde{z}_1=\delta_1(z_1-x_1),\ \tilde{z}_2=\delta_2(z_2-x_2), \ \tilde{\xi}_1=\delta_1 \zeta_1+\frac{1}{2}(a+c)\eta_1-\frac{1}{2}c\eta_2,$$
$$\tilde{\xi}_2=\delta_2 \zeta_2-\frac{1}{2}c\eta_1+\frac{1}{2}(b+c)\eta_2, \ \tilde{\eta}_1=-\frac{1}{2}\xi_1, \ \tilde{\eta}_2=-\frac{1}{2}\xi_2, \ \tilde{\zeta}_1=-\delta_1 \zeta_1-\frac{1}{2}\eta_1,$$
$$\tilde{\zeta}_2=-\delta_2 \zeta_2-\frac{1}{2}\eta_2.$$
It follows that 
$$\textrm{Ker}(\textrm{Re }F) \cap \rr^{12}=\{(x,y,z,\xi,\eta,\zeta) \in \rr^{12} : \zeta=0, \ x=z\},$$
$$\textrm{Ker}(\textrm{Re }F) \cap \textrm{Ker}(\textrm{Re }F\textrm{Im }F) \cap \rr^{12}=\{(x,y,z,\xi,\eta,\zeta) \in \rr^{12} : y=\eta=\zeta=0, \ x=z\}$$
and
\begin{multline*}
\textrm{Ker}(\textrm{Re }F) \cap \textrm{Ker}(\textrm{Re }F\textrm{Im }F) \cap \textrm{Ker}\big(\textrm{Re }F(\textrm{Im }F)^2\big) \cap \rr^{12}\\ =\{ y=\xi=\eta=\zeta=0, \ x=z, \ (a+c-1)x_1-cx_2=0, -cx_1+(b+c-1)x_2=0\}.
\end{multline*}
When the condition 
\begin{equation}\label{osc4}
(a+c-1)(b+c-1)-c^2 \neq 0,
\end{equation}
holds, the singular space reduces to $\{0\}$ after intersecting exactly $k_0+1$ kernels with here $0 \leq k_0=2 \leq 11$. Notice that condition (\ref{osc4}) corresponds exactly to the assumption $V(x)-x^2/2$ is a Morse function required in~\cite{HeHiSj2} (Lemma~6.1) to ensure that the needed dynamical conditions hold. The other conditions
$\partial_x^{\alpha}V_j(x)=\mathcal{O}(1)$, when $|\alpha| \geq 2$, for $j=1,2,c$ and $|\nabla_xV(x)-x| \geq 1/C$ when $|x| \geq C$, are also satisfied for quadratic potentials fulfilling condition (\ref{osc4}). When this condition holds, Theorems~\ref{th1} and~\ref{th2} applies. This shows in particular that the first eigenvalue\footnote{which may be computed explicitly by using the formula (\ref{vp0})} in the bottom of the spectrum  $\mu_0$ has algebraic multiplicity one with an eigenspace 
\begin{equation}\label{ji0}
\textrm{Ker}\big(q^w(X,D_X)-\mu_0\big)=\cc \mathcal{M}_{\alpha_1,\alpha_2},
\end{equation}
spanned by a ground state of exponential type 
\begin{equation}\label{ji0.5}
\mathcal{M}_{\alpha_1,\alpha_2}(x,y,z)=e^{-a(x,y,z)} \in \mathscr{S}(\rr^6),
\end{equation}
where $a$ is a complex-valued quadratic form on $\rr^6$ whose real part is positive definite. Notice that the operator 
\begin{multline*}
q^w(X,D_X)u=2u+\alpha_1 \Big(-\partial_{z_1}+\frac{1}{\alpha}(z_1-x_1)\Big)\Big(\partial_{z_1}+\big(\frac{2}{\alpha_1}-\frac{1}{\alpha}\Big)(z_1-x_1)\Big)u+\\ \alpha_2\Big(-\partial_{z_2}+\frac{1}{\alpha}(z_2-x_2)\Big)\Big(\partial_{z_2}+\Big(\frac{2}{\alpha_2}-\frac{1}{\alpha}\Big)(z_2-x_2)\Big)u+\Big(y\partial_x-(\nabla_xV(x)-z)\partial_y\Big)u,
\end{multline*}
is real. The discussion preceding the statement of Theorem~\ref{th3} then shows that the quadratic form $a$ is positive definite. This proves the existence of a Maxwellian $\mathcal{M}_{\alpha_1,\alpha_2}$ in the general case when the temperatures may be different $\alpha_1 \neq \alpha_2$. It would be interesting to push further the computations in order to derive the exact expressions for the eigenvalue $\mu_0$, the spectral gap $\tau_0$ and the Maxwellian $\mathcal{M}_{\alpha_1,\alpha_2}$. Another question of interest would be to investigate if Theorem~\ref{th3} applies by checking the validity of condition (\ref{y11}). Those questions are left for a future work. From now, we may only notice from (\ref{y1}) and (\ref{dl3}) that the following resolvent estimate holds 
$$\exists c>0,  \ \{z \in \cc : \textrm{Re }z \geq -1/2, \ \textrm{Re }z +1 \leq c|z+1|^{\frac{1}{5}}\} \cap \sigma(q^w(X,D_X))=\emptyset,$$
$$\exists C>0, \ \|(q^w(X,D_X)-z)^{-1}\| \leq C |z+1|^{-\frac{1}{5}},$$
for all $z \in \cc$ such that $\textrm{Re }z \geq -1/2$ and $\textrm{Re }z +1 \leq c|z+1|^{\frac{1}{5}}$, since the integer $k_0$ is here equal to 2.


\end{document}